\documentclass[aps,prl,reprint,amsmath,amssymb,showpacs,superscriptaddress]{revtex4-1}
\usepackage{graphicx}
\usepackage{dcolumn}
\usepackage{bm}
\usepackage{color}
\usepackage{hyperref}

\begin{document}
\title{Chiral symmetry and peripheral neutron-$\alpha$ scattering}
\author{Yilong Yang}
\affiliation{State Key Laboratory of Nuclear Physics and Technology, School of Physics, Peking University, Beijing 100871, China}

\author{Evgeny Epelbaum}
\email{evgeny.epelbaum@ruhr-uni-bochum.de}
\affiliation{Ruhr-Universit\"at Bochum, Fakult\"at f\"ur Physik und Astronomie, Institut f\"ur Theoretische Physik II, D-44780 Bochum, Germany}

\author{Jie Meng}
\email{mengj@pku.edu.cn}
\affiliation{State Key Laboratory of Nuclear Physics and Technology, School of Physics, Peking University, Beijing 100871, China}

\author{Lu Meng}
\affiliation{Ruhr-Universit\"at Bochum, Fakult\"at f\"ur Physik und Astronomie, Institut f\"ur Theoretische Physik II, D-44780 Bochum, Germany}
\affiliation{School of Physics, Southeast University, Nanjing 211189, China}

\author{Pengwei Zhao}
\email{pwzhao@pku.edu.cn}
\affiliation{State Key Laboratory of Nuclear Physics and Technology, School of Physics, Peking University, Beijing 100871, China}

\begin{abstract}
We propose and demonstrate that peripheral neutron-$\alpha$ scattering at low energies can serve as a sensitive and clean probe of the long-range three-nucleon forces. To this aim, we perform  {\it ab initio} quantum Monte Carlo calculations using two- and three-nucleon interactions derived in chiral effective field theory up to third expansion order. We show that the longest-range three-nucleon force stemming from the two-pion exchange plays a crucial role in the proper description of the neutron-$\alpha$ $D$-wave phase shifts.
Our Letter reveals the predictive power of chiral symmetry in the few-body sector  and opens a new direction for probing and constraining three-nucleon forces.
\end{abstract}

\maketitle
Chiral effective field theory (EFT)~\cite{Weinberg1990Phys.Lett.B288, Weinberg1991Nucl.Phys.B318} provides a solid foundation for understanding the interactions between protons and neutrons, one of the fundamental problems in physics. The method relies on the spontaneously broken approximate chiral symmetry of QCD, which allows one to connect nuclear forces with the underlying theory of the strong interactions between quarks and gluons \cite{Gross:2022hyw} in a theoretically clean way \cite{Epelbaum2009Rev.Mod.Phys.1773,Machleidt2011Phys.Rep.175,Hammer2020Rev.Mod.Phys.025004}.
In particular, the long-range behavior of nuclear interactions and current operators is determined in a parameter-free manner by the chiral symmetry (and its breaking pattern), along with the experimental information on the pion-nucleon ($\pi$N) system required to pin down the corresponding low-energy constants (LECs) \cite{Epelbaum:2024gfg}.

In the two-nucleon (NN) sector, peripheral neutron-proton scattering has been traditionally used to probe the long-range NN interaction and test the convergence and predictive power of chiral EFT \cite{Kaiser1997Nucl.Phys.A758,Entem2002Phys.Rev.C014002,Birse:2003nz}. However, the large-distance
behavior of the NN force is strongly dominated by the one-pion exchange (OPE), whose form is in fact not sensitive to the chiral symmetry: indeed, both the pseudovector $\pi$N coupling dictated by the chiral symmetry and the chiral-symmetry-violating  pseudoscalar one lead to the same on-shell OPE potential.
In contrast, the two-pion exchange (TPE) is strongly constrained by the chiral symmetry and comes out as a nontrivial prediction of chiral EFT. In spite of the large-distance dominance of the OPE, clear evidence of the chiral TPE was observed in NN scattering data using the last generation of high-precision chiral NN potentials \cite{Epelbaum2015Phys.Rev.Lett.122301,Reinert2018Eur.Phys.J.A86}.

Compared to the NN interaction, three-nucleon forces (3NFs) are not yet well understood and constitute an important frontier in low-energy nuclear physics \cite{Kalantar-Nayestanaki:2011rzs,Hammer:2012id,Endo:2024cbz}.
The dominant 3NF contributions at third order (N$^2$LO) in the chiral expansion are well established \cite{vanKolck1994Phys.Rev.C2932,Epelbaum2002Phys.Rev.C064001} and have been widely used in {\it ab initio} calculations of nuclear structure and reactions; see Ref.~\cite{Endo:2024cbz} and the references therein. Higher-order corrections are currently being derived using symmetry-preserving cutoff regularization \cite{Krebs:2023ljo,Krebs:2023gge}. Similarly to the NN force, the long- and intermediate-range behavior of the 3NF is predicted by the chiral symmetry of QCD. But how can these predictions be tested experimentally?

\begin{figure}[htb]
    \centering
    \includegraphics[width=0.99\linewidth]{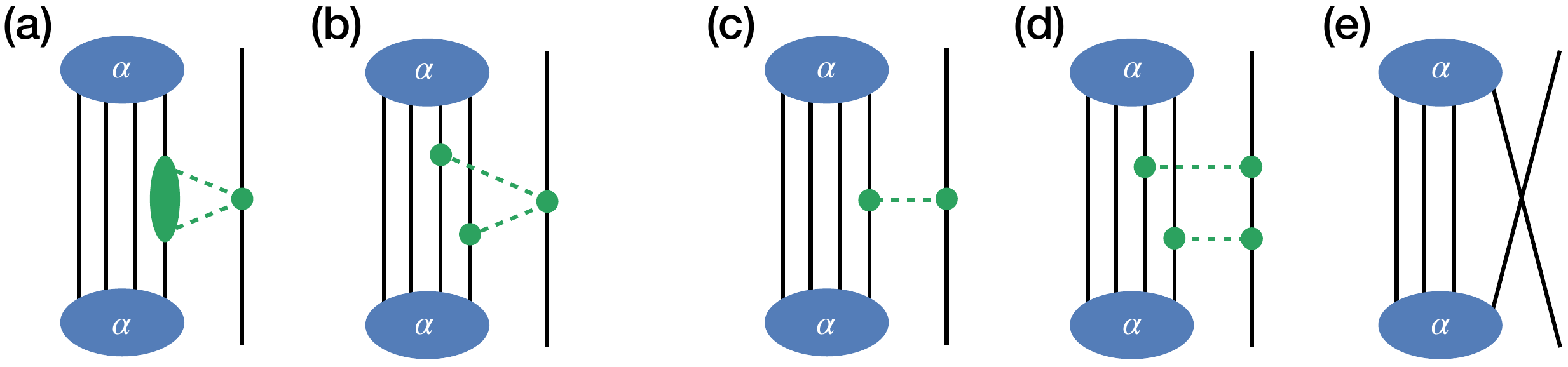}
    \caption{Diagrams (a) and (b) show the dominant contributions to low-energy peripheral neutron-$\alpha$ scattering. The OPE process (c) is forbidden by the isospin selection rule for the isospin $T=0$ $\alpha$-particle. Iterated OPE (d) is suppressed in the tight-binding limit while short-range interactions (e) are suppressed in peripheral scattering by the centrifugal barrier. Solid and dashed lines denote nucleons and pions, respectively, while green-shaded ellipses denote the corresponding $\pi$N amplitudes.
    }
    \label{fig1}
\end{figure}

Already in Refs.~\cite{Barreiro1998Phys.Rev.C21422148, Higa1999arXiv9908062}, it was speculated that peripheral neutron-$\alpha$ scattering might be used to probe the long-range tail of the 3NF.
This process is advantageous compared to other few-nucleon reactions due to the high inelastic threshold and the isoscalar nature of the $\alpha$-particle that suppresses the OPE \cite{Barreiro1998Phys.Rev.C21422148, Higa1999arXiv9908062}, see Fig.~\ref{fig1}.
However, an \textit{ab initio} description of five-body scattering is a challenging task from a numerical perspective. The first \textit{ab initio} study of neutron-$\alpha$ scattering using realistic interactions was performed within the quantum Monte Carlo (QMC) approach~\cite{Carlson1987Phys.Rev.C2731}; see Refs.~\cite{Nollett2007Phys.Rev.Lett.022502, Lynn2016Phys.Rev.Lett.062501, Flores2025PRC112.014008} for more recent results.
Later, neutron-$\alpha$ scattering was also studied within the Faddeev-Yakubovsky framework~\cite{Lazauskas2018Phys.Rev.C044002, Lazauskas2020Front.Phys.}, the no-core shell model methods~\cite{Kravvaris2020Phys.Rev.C024616, Shirokov2018Phys.Rev.C044624,   Navratil2016Phys.Scr.053002, Zhang2020Phys.Rev.Lett.112503, Mercenne2022Comput.Phys.Commun.108476}, the stochastic variational approach~\cite{Bagnarol2023Phys.Lett.B138078}, and the nuclear lattice EFT~\cite{elhatisari2025arXiv}.
However, the existing studies are limited to $S$ and $P$ partial waves dominated by short-range mechanisms, while no \textit{ab initio} studies of peripheral neutron-$\alpha$ scattering are available yet to probe the long-range tail of the 3NF.

In this Letter we fill this gap and demonstrate that peripheral neutron-$\alpha$ scattering below the inelastic threshold provides a clean and sensitive testing ground for the longest-range TPE component of the 3NF, thereby allowing one to directly probe the implications of the chiral symmetry in the few-nucleon sector. To this aim,
we present QMC calculations of neutron-$\alpha$ $D$-wave phase shifts using the nuclear Hamiltonian constructed in chiral EFT at the three lowest expansion orders $Q^0$ (LO), $Q^2$ (NLO) and $Q^3$. Here and in what follows, $Q \in \{M_\pi/\Lambda_b, \; |\vec p \, |/\Lambda_b\}$ denotes the expansion parameter in chiral EFT with $M_\pi$, $\vec p$ and $\Lambda_b$  referring to the pion mass, a typical nucleon momentum and the breakdown scale, respectively. Our results reveal a significant impact of the TPE 3NF on the neutron-$\alpha$ $D$-wave phase shifts.
A comparison with the empirical phase shifts from the $R$-matrix analysis of the experimental data allows one to
verify and quantify the role of the chiral symmetry in constraining the 3NF.

Throughout this Letter we use local chiral EFT interactions which are well suited for QMC applications.
The employed  Hamiltonian has the form
\begin{equation}\label{eq.H}
    H=\sum_{i=1}^A\frac{-\nabla_i^2}{2m_N}+\sum^A_{i<j}v_{ij}+\sum^A_{i<j<k}V_{ijk},
  \end{equation}
with $v_{ij}$ and $V_{ijk}$ denoting the local versions of the NN potential \cite{Gezerlis2013Phys.Rev.Lett.032501, Gezerlis2014Phys.Rev.C054323} and the 3NF \cite{Tews2016Phys.Rev.C024305, Lynn2016Phys.Rev.Lett.062501, Lynn2017Phys.Rev.C054007}, respectively. 
We employ here the cutoff value of $R_0=1.2$~fm, which corresponds to a typical momentum cutoff $\Lambda\sim400$~MeV.
Pion-exchange contributions to $v_{ij}$ and $V_{ijk}$ are regulated by a smooth cutoff at short distances $r<R_0$, leaving their long-distance behavior predicted by chiral EFT largely unaffected.
The Coulomb force and isospin-breaking NN interactions are neglected since they are expected to have a small effect on neutron-$\alpha$ scattering.

The 3NF at N$^2$LO takes the form
\begin{equation}
    V=V_{2\pi}+V_{1\pi\rm -cont}+V_{\rm cont}\,.
  \end{equation}
  The longest-range TPE term $V_{2\pi}$ depends on the $\pi$N LECs $c_1$, $c_3$ and $c_4$, for which we adopt the values of $c_1=-0.81$ GeV$^{-1}$, $c_3=-3.40$ GeV$^{-1}$, and $c_4=3.40$ GeV$^{-1}$~\cite{Epelbaum2005Nucl.Phys.A362} from the order-$Q^3$ analysis of the $\pi N$ system~\cite{Buettiker2000Nucl.Phys.A97}, which are consistent with those used in the NN potential of Ref.~\cite{Gezerlis2014Phys.Rev.C054323}.
Further, $V_{1\pi\rm -cont}$ and $V_{\rm cont}$ denote the short-range components of the 3NF [see Eqs.~(2b) and (5a) of Ref. \cite{Lynn2016Phys.Rev.Lett.062501}]. For the corresponding LECs $c_D$ and $c_E$, we employ the values $c_D=3.5$ and $c_E=0.09$ from Ref.~\cite{Lynn2016Phys.Rev.Lett.062501}, which were fitted to the $\alpha$ binding energy and the low-energy neutron-$\alpha$ $P$-wave phase shifts.

To perform QMC calculations of neutron-$\alpha$ phase shifts, we first convert the scattering problem to an eigenvalue problem by placing the nucleons in a harmonic oscillator (HO) trap. This is achieved by adding the NN potential
\begin{equation}
    V_{\rm HO}=\sum_{i<j}\frac{1}{2}\frac{m_N}{A}\omega^2 r_{ij}^2,
\end{equation}
where $m_N$ is the nucleon mass, $\omega$ the HO frequency and $r_{ij}$ the distance between two nucleons.
Then, the BERW formula~\cite{Busch1998FoundationsofPhysics549, Suzuki2009Phys.Rev.A033601} is used to extract the neutron-$\alpha$ $D$-wave phase shifts from the calculated $^5$He eigenstate energies via
\begin{equation}\label{eq.BERW} 
\tan\delta_{n\alpha}(E_{J^\pi})=-\left(\frac{E_{J^\pi}}{2\omega}\right)^{\frac{5}{2}}\frac{\Gamma\left(-\frac{3}{4}-\frac{E_{J^\pi}}{2\omega}\right)}{\Gamma\left(\frac{7}{4}-\frac{E_{J^\pi}}{2\omega}\right)}.
\end{equation}
Here, $E_{J^\pi}=E(^5{\rm He}_{J^\pi})-E(^4{\rm He}_{0^+})$ denotes the energy of the trapped $^5$He eigenstate with respect to the neutron-$\alpha$ energy threshold given by the (trapped) $\alpha$ ground-state energy.
In QMC calculations, the energy of $^5$He ($^4$He) is calculated directly in the center-of-mass frame to remove spurious center-of-mass motions, by subtracting the center-of-mass position from all the spatial coordinates of nucleons, $\bm r_i\rightarrow \bm r_i-\bm R_{\rm cm}$~\cite{Massella2020J.Phys.G47.035105}.
There are two $D$-wave scattering channels $^2D_{\frac{5}{2}}$ and $^2D_{\frac{3}{2}}$, which correspond to the $^5$He eigenstates with quantum numbers $J^\pi=\frac{5}{2}^+$ and $J^\pi=\frac{3}{2}^+$, respectively.
Owing to the fact that the spin-orbit splitting between the two channels is negligible at low energies~\cite{Bond1977Nucl.Phys.A317},
we only consider here the $^2D_{\frac{5}{2}}$ channel.

The ground state energy of $^5$He $(J^\pi=\frac{5}{2}^+)$ is calculated using the Green's function Monte Carlo (GFMC) method with newly developed neural-network wave functions.
So far, most of the existing neural-network wave functions of nuclei~\cite{Adams2021Phys.Rev.Lett.022502, Gnech2021FewBodySystems7, Yang2022Phys.Lett.B137587, Fore2023Phys.Rev.Res.033062, Gnech2024Phys.Rev.Lett.142501} are designed for variational Monte Carlo rather than GFMC calculations.
Only recently, the neural-network representation of the wave function suitable for GFMC calculations of $A\leq4$ nuclei with realistic NN interactions has been developed \cite{Yang2025ChinesePhys.Lett.051201}.
In this Letter, we improve the neural-network wave function in Ref.~\cite{Yang2025ChinesePhys.Lett.051201} by considering a generalized backflow transformation~\cite{Luo2019Phys.Rev.Lett.226401, Yang2023Phys.Rev.C034320}, and adapt it to neutron-$\alpha$ scattering calculations.

The neural-network wave function takes the form
\begin{equation}
    |\Psi_T\rangle_{J^\pi}=\hat{\mathcal F}\mathcal{A}|\uparrow_n\downarrow_n\uparrow_p\downarrow_p \phi^{J^\pi}_n\rangle,
\end{equation}
where the first four single-particle states represent the four $s$-state nucleons in the $\alpha$ cluster, while the last single-particle orbital $\phi^{J^\pi}_n$ corresponds to the neutron scattered off the $\alpha$ cluster, thus taken to be the HO solution with quantum number $J^\pi$.
This enforces the correct asymptotic behavior of the wave function at large neutron-$\alpha$ separation in the HO trap.
In the above equation, $\hat{\mathcal F}$ is a neural-network correlator of the following form,
\begin{equation}\label{eq.FR}
    \hat{\mathcal{F}}(\bm R)={\rm e}^{\mathcal{U}(\bm R)}\tanh\mathcal{V}(\bm R)\left(
    1 + \sum_{i<j}^A\sum_{p=2}^6 \mathcal{W}_{ij}^p(\bm R)\hat{O}_{ij}^p
    \right).
\end{equation}
Here, $\bm R=(\bm r_1,\bm r_2,\ldots,\bm r_A)$ and $\hat{O}_{ij}^{p=2\text{-}6}$ are two-body spin-isospin operators, $\hat{O}_{ij}^{p=2\text{-}6}=\bm\tau_i\cdot\bm\tau_j, \bm\sigma_i\cdot\bm\sigma_j, \bm\sigma_i\cdot\bm\sigma_j\bm\tau_i\cdot\bm\tau_j,  S_{ij}, S_{ij}\bm\tau_i\cdot\bm\tau_j$,
with $S_{ij}=3\bm\sigma_i\cdot\bm r_{ij}\bm\sigma_j\cdot\bm r_{ij}-\bm\sigma_i\cdot\bm\sigma_j r_{ij}^2$.
The central correlation functions $\mathcal{U},\mathcal{V}$ are represented by a permutation-invariant neural network~\cite{Adams2021Phys.Rev.Lett.022502, Gnech2021FewBodySystems7}.
The distances of nucleon pairs are used as input to ensure the translational and rotational invariance of the trial wave function~\cite{Yang2022Phys.Lett.B137587}.
The spin-isospin dependent correlation function $\mathcal{W}$ has the form
\begin{equation}
    \mathcal{W}_{ij}^p(\bm R)=\rho_{\mathcal{W}}^p\left(\sum_{k=1}^A\phi_\mathcal{W}(r_{ij}, r_{ik}, r_{jk})\right).
\end{equation}
Here, the generalized backflow transformation is implemented to take into account correlation effects~\cite{Yang2023Phys.Rev.C034320}, which has been shown to augment the performance of neural-network wave functions for nuclei~\cite{Fore2023Phys.Rev.Res.033062, Gnech2024Phys.Rev.Lett.142501}.

The ``conventional" way of introducing correlations in the wave function relies on products of two- and three-body functions of some assumed kind~\cite{Wiringa2000Phys.Rev.C014001}. In contrast, the neural network can more effectively represent the correlation functions in a compact form, thanks to the universal approximation theorem~\cite{Hornik1989NeuralNetworks359}. Given the same ansatz [Eq.~\eqref{eq.FR}], the neural-network correlation functions lead to significantly improved variational energies~\cite{Adams2021Phys.Rev.Lett.022502,Gnech2021FewBodySystems7,Yang2022Phys.Lett.B137587,Yang2023Phys.Rev.C034320} and, thus, provide better initial states for the GFMC calculations~\cite{Yang2025ChinesePhys.Lett.051201}.

Starting from the trained neural-network wave function, we carry out the GFMC calculations to project out the lowest $J^\pi=\frac{5}{2}^+$ eigenstate via the imaginary time propagation~\cite{Carlson1987Phys.Rev.C2731},
\begin{equation}
    \lim_{\tau\rightarrow\infty}{\rm e}^{-H\tau}|\Psi_T\rangle_{J^\pi}\rightarrow|\Psi_0\rangle_{J^\pi}.
\end{equation}
This is carried out by a sequence of short-time propagation ${\rm e}^{-H\Delta\tau}$ using a branching random walk algorithm with importance sampling~\cite{Pudliner1997Phys.Rev.C1720, Wiringa2000Phys.Rev.C014001}.
When performing the GFMC propagation, one has to deal with the sign problem that causes large statistical fluctuations at large $\tau$.
Following the previous GFMC calculations of neutron-$\alpha$ scattering~\cite{Nollett2007Phys.Rev.Lett.022502}, we use a transient estimate to mitigate the sign problem.
We first perform the constrained-path propagation~\cite{Wiringa2000Phys.Rev.C014001}, which suppresses the sign problem, and then release the constraints to obtain the final result.
Such transient estimates result in significantly improved estimates compared to those without performing constrained-path propagation at first.
Moreover, the softest available cutoff value of $R_0=1.2$ fm is employed in the chiral nuclear Hamiltonian~(\ref{eq.H}) to avoid as much as possible the sign problem.
See the Supplemental Material~\cite{Supp} for more details on the present GFMC calculations with neural-network wave functions.

\begin{figure}[!htbp]
    \centering
    \includegraphics[width=0.9\linewidth]{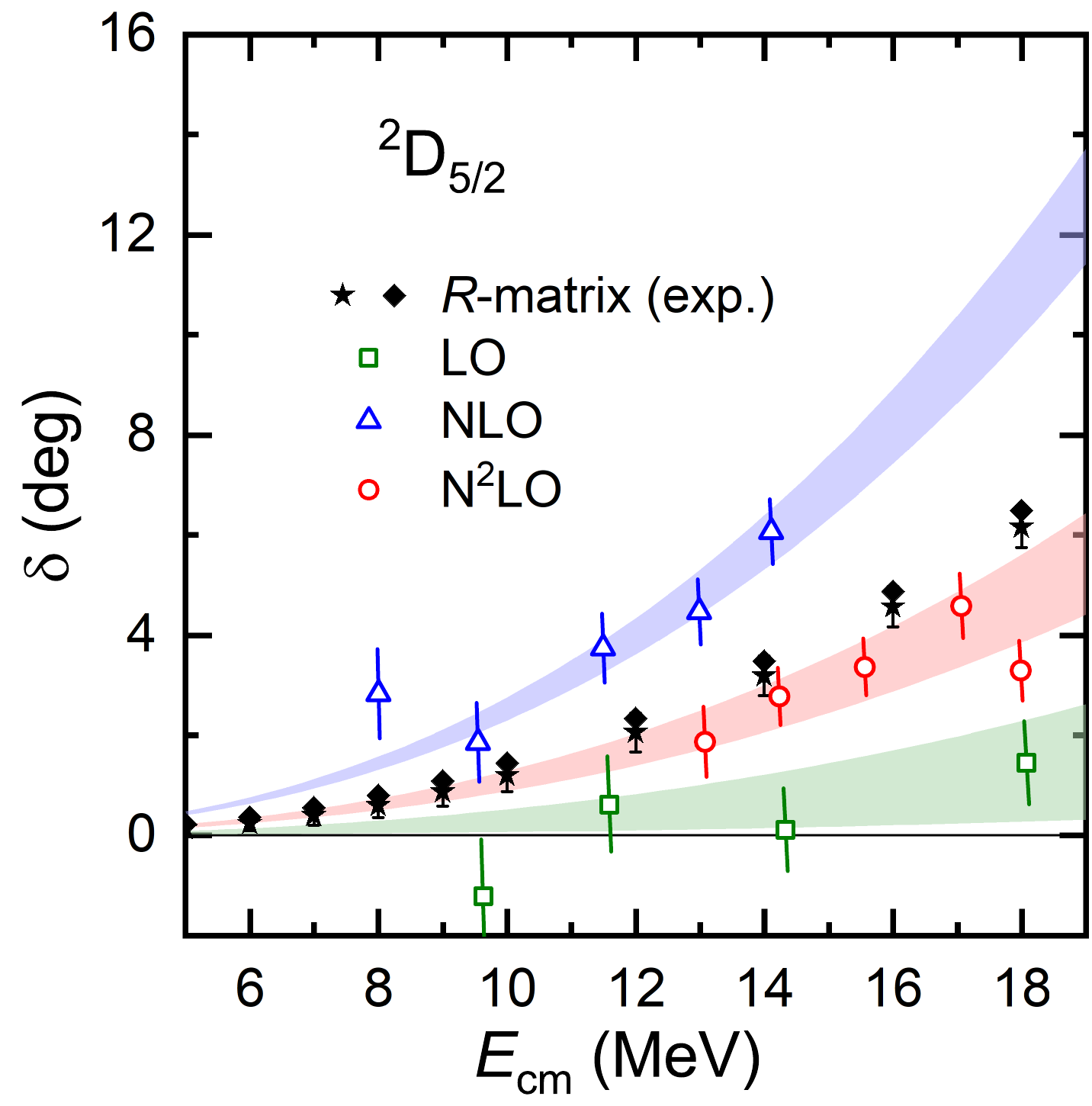}
    \caption{(Color online). Phase shifts for neutron-$\alpha$ scattering in the $^2D_{\frac{5}{2}}$ channel as a function of the center-of-mass energy predicted at different orders of the chiral EFT expansion.
    Empty symbols refer to the GFMC results with error bars denoting the statistical uncertainties.
    The $1\sigma$ uncertainty bands come from analyzing the GFMC results, which combine the statistical uncertainties and the systematic uncertainties of the BERW formula (see Supplemental Material~\cite{Supp} for details).
    The stars and diamonds are from the $R$-matrix analyses of the experimental neutron-$\alpha$ elastic scattering data from~\cite{Bond1977Nucl.Phys.A317} and~\cite{Hale2025}, respectively.
    }
    \label{fig2}
\end{figure}

Figure \ref{fig2} depicts the $D$-wave neutron-$\alpha$ phase shift predicted at different orders of chiral EFT as a function of the center-of-mass energy $E_{\rm c.m.}$.
Each point in the figure corresponds to a single GFMC calculation using a HO trap, and it is derived from the BERW formula [Eq.~(\ref{eq.BERW})].
The values of HO frequency $\omega$ are chosen such that the oscillator lengths $b=\sqrt{2/(m_N\omega)}$ are in the range of $b=4\text{--}6$ fm, which is much larger than the NN interaction range and the size of the $\alpha$ particle. 
Here, the oscillator length $b$ is defined for NN relative motion.
Moreover, the chosen values of $\omega$ are small such that the effect of the HO trap on the $\alpha$ cluster's internal structure is about an order less than the energy of inelastic $\alpha$ excitation. For the case of $D$-wave NN scattering, we have verified the accuracy of the BERW formula for extracting phase shifts with HO traps with $b=4\text{--}6$~fm to be within $\sim10\%$~\cite{Supp}.

We implement the theoretically well-founded EFT-based formalism~\cite{Zhang2020Phys.Rev.C051602} to quantify the systematic errors of the BERW formula for extracting phase shifts; see also Refs.~\cite{Luu2010Phys.Rev.C82.034003, Zhang2020Phys.Rev.Lett.112503, Li2021Phys.Rev.C104.044001} on this topic. In this formalism, the systematic errors from finite HO frequency are accounted for by an expansion in terms of $\mu\omega/M_H^2$, with $\mu=4m_N/5$ the reduced mass and $M_H\sim200$ MeV a high-momentum breakdown scale. We merge this formalism with Bayesian methodology~\cite{Zhang2020Phys.Rev.Lett.112503} to determine the EFT parameters from the limited number of GFMC calculations at hand and to combine the systematic and statistical errors of the phase shifts; see the Supplemental Material~\cite{Supp} for details on the uncertainty quantification.

We compare the predicted $D$-wave neutron-$\alpha$ phase shifts to those obtained from the $R$-matrix analyses of experimental neutron-$\alpha$ elastic scattering data~\cite{Bond1977Nucl.Phys.A317, Hale2025}.
The LO chiral Hamiltonian involves only the OPE and derivativeless short-range NN interactions and yields nearly vanishing $D$-wave neutron-$\alpha$ phase shifts in line with the arguments provided in the introduction; see Fig.~\ref{fig1}.
The dominant contribution to the phase shift is generated at NLO when the chiral TPE NN force enters, but it significantly overestimates the $R$-matrix phase shifts.
Then, the N$^2$LO corrections considerably reduce the predicted phase shifts and bring the theoretical results closer to the $R$-matrix phase shifts.

\begin{figure}[!htbp]
    \centering
    \includegraphics[width=0.9\linewidth]{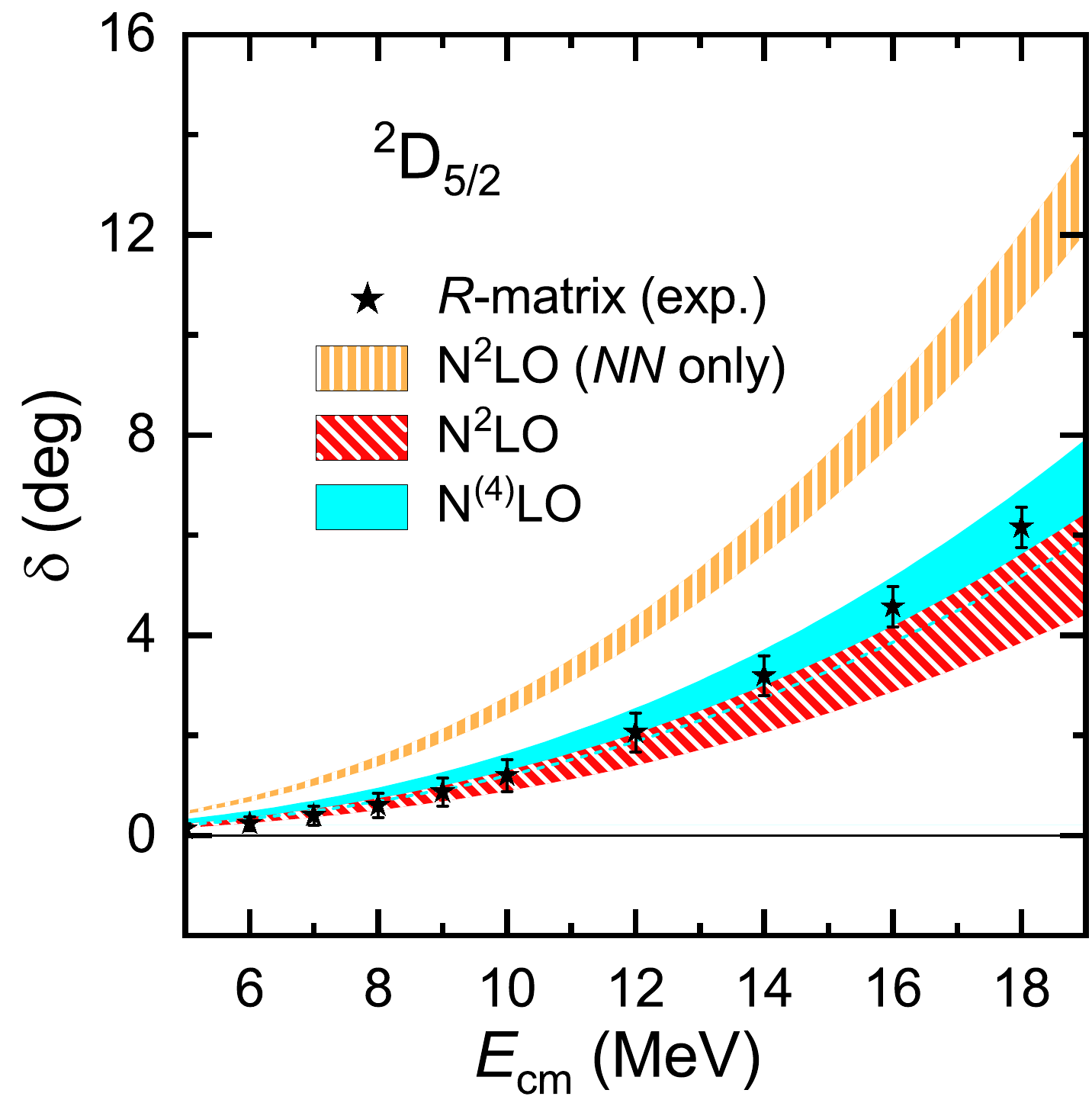}
    \caption{Phase shifts for neutron-$\alpha$ scattering in the $^2D_{\frac{5}{2}}$ channel as a function of the center-of-mass energy obtained at N$^2$LO with and without the 3NF.
    The N$^{(4)}$LO predictions show the results based on the effective order-$Q^5$ values of the $\pi$N LECs in the 3NF as explained in the text.
    The stars are from the $R$-matrix analysis of experimental data~\cite{Bond1977Nucl.Phys.A317}.
    }
    \label{fig3}
\end{figure}

The N$^2$LO corrections to the Hamiltonian consist of the subleading TPE NN potential and the leading contribution to the 3NF. To disentangle the impact of the two- and three-nucleon interactions, we show in Fig.~\ref{fig3} the N$^2$LO predictions obtained with and without the inclusion of the 3NF.
These results reveal that the subleading TPE NN potential plays a minor role, as one would expect for the considered soft value of the regulator $R_0=1.2$ fm. In contrast,  the 3NF plays a key role in bringing the calculated phase shift in a better agreement with the empirical results. Moreover, for a harder regulator $R_0=1.0$ fm, the 3NF also has a significant impact on the phase shifts.
This clearly demonstrates that $D$-wave neutron-$\alpha$ scattering provides a very sensitive tool to probe the long-distance behavior of the 3NF.

Compared to the analysis of a folding model in Ref.~\cite{Higa1999arXiv9908062}, the present results find a similar overestimation of $D$-wave phase shifts with chiral NN forces only. However, the TPE 3NF only slightly reduces the phase shifts by 5\% in the folding model, while it has significant contributions in the present calculations. Such difference highlights the importance of the present \textit{ab initio} calculations that respect the antisymmetric nature of few-nucleon states.

The N$^2$LO prediction of the phase shifts of the $D$ waves appears to slightly underestimate the empirical results at high energies, which means that the repulsive contribution of the 3NF is slightly too large. This can possibly be attributed to the known overestimation of the longest-range TPE 3NF components at this chiral order. In Refs.~\cite{Bernard:2007sp,Krebs2012Phys.Rev.C054006}, the order-$Q^4$ (N$^3$LO) and order-$Q^5$ (N$^4$LO) corrections to the TPE 3NF have been worked out and shown to reduce the strength of the dominant N$^2$LO contribution. It was found in Ref.~\cite{Krebs2012Phys.Rev.C054006} that using the N$^2$LO expressions for the TPE 3NF with the smaller-in-magnitude values of the $\pi$N LECs $c_1=-0.37$ GeV$^{-1}$, $c_3=-2.71$ GeV$^{-1}$ and $c_4=1.41$ GeV$^{-1}$ allows one to effectively take into account corrections to this 3NF topology up through N$^4$LO. Here, we confirm these findings by demonstrating that the effective order-$Q^5$ values of $c_i$'s in the 3NF indeed lead to an improved description of the neutron-$\alpha$ $D$-wave phase shift; see Fig.~\ref{fig3}.

\begin{figure}[!htbp]
    \centering
    \includegraphics[width=0.9\linewidth]{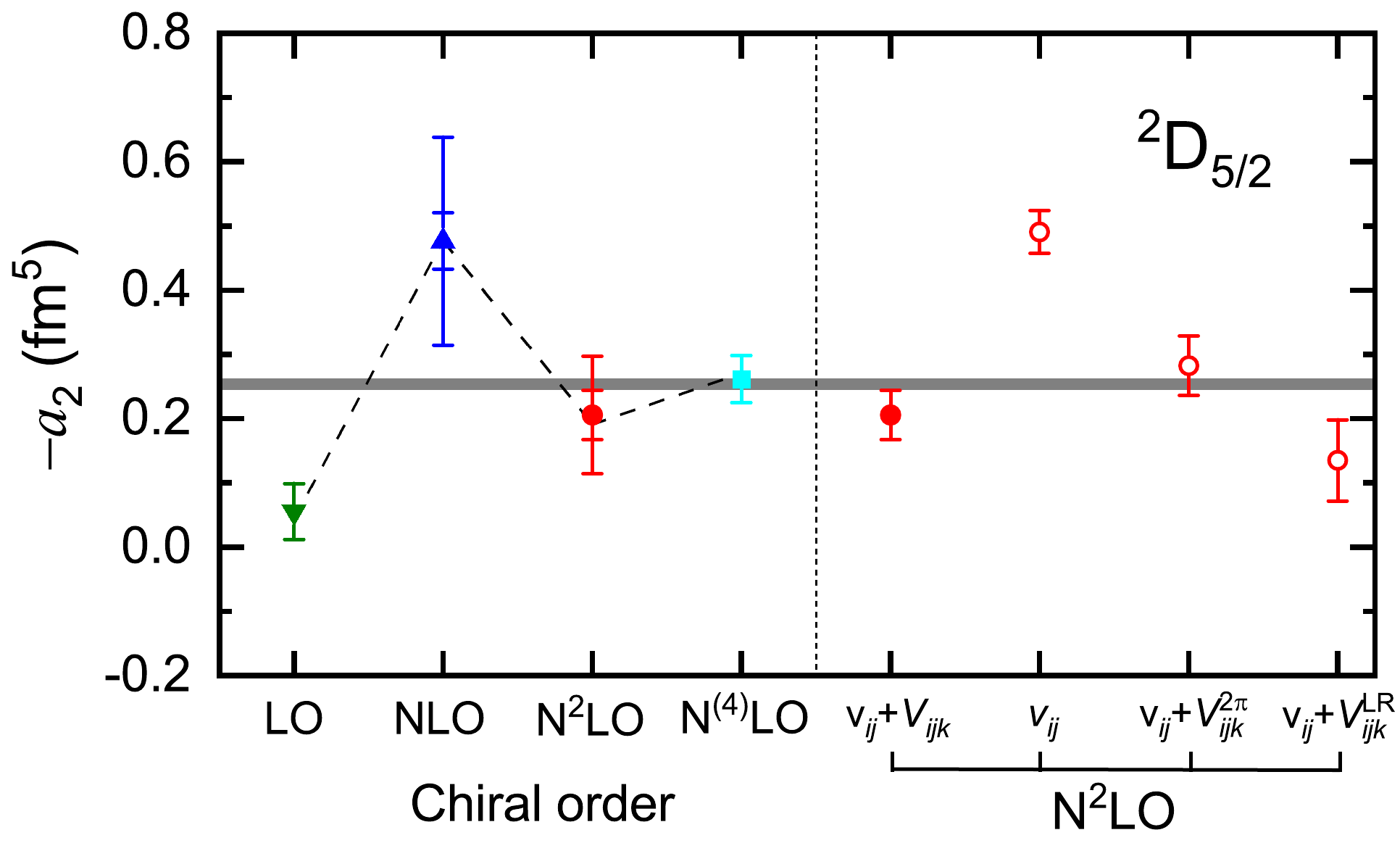}
    \caption{Left panel: chiral EFT predictions for $D$-wave scattering lengths $a_2$ at different expansion orders.
    Smaller error bars indicate numerical uncertainties, while larger error bars are uncertainties from the truncation of the chiral expansion.
    The right panel shows the results obtained at N$^2$LO with the full NN$+$3N forces ($v_{ij}+V_{ijk}$), the NN force only ($v_{ij}$), the NN force and the TPE 3NF ($v_{ij}+V_{ijk}^{2\pi}$), and the NN force and the long-range 3NF ($v_{ij}+V_{ijk}^{\rm LR}$).
    Here, $V_{ijk}^{\rm LR}$ is defined by switching off the 3NF $V$ at nucleons' relative distances $r < 1.5$~fm.
    The horizontal black band shows the $a_2$ value from the $R$-matrix analysis \cite{Bond1977Nucl.Phys.A317}.
    }
    \label{fig4}
\end{figure}
The smallness of the phase shifts suggests that scattering in the $^2D_{\frac{5}{2}}$ channel is perturbative and the phase shifts can, to a good approximation, be described by
\begin{equation}\label{eq.a2}
    \tan\delta_{n\alpha}(k)=-a_2 k^5+O(k^7),
\end{equation}
where $a_2$ is the scattering ``length" and $k=\sqrt{2\mu E_{\rm c.m.}}$ the neutron-$\alpha$ relative momentum with $\mu=4m_N/5$ being the reduced mass.

Finally, we show in the left panel of Fig.~\ref{fig4} the predicted $D$-wave scattering length $a_2$ at different EFT orders.
The theoretical uncertainties from the truncation of the chiral expansion are estimated using the Bayesian model $\bar C_{0.5-10}^{650}$ \cite{Epelbaum2020EPJA92}.
We find that the N$^{(4)}$LO prediction agrees well with the $R$-matrix result and also with the N$^2$LO prediction (within uncertainty).

The impact of the individual N$^2$LO contributions to the Hamiltonian on the scattering length $a_2$ is shown in the right panel of Fig.~\ref{fig4}.
The value using the NN force alone is significantly larger than both the complete N$^2$LO prediction and the result from the $R$-matrix analysis.
With only the TPE 3NF $V_{2\pi}$, the overestimation of $a_2$ from the NN force alone can already be resolved.
In addition, the $a_2$ value obtained with the full 3NF $V$ and that obtained with the long-range 3NF $V_{\rm LR}$, defined by switching off the 3NF at the relative distances of the nucleons with $r < 1.5$~fm, are
consistent with each other within uncertainties.
This suggests that, as expected, short-range 3NF contributions have minor effects on $a_2$.

In summary, we have performed the first \textit{ab initio} study of neutron-$\alpha$ $D$-wave scattering in the framework of chiral EFT. The phase shifts have been extracted from the QMC calculations of the $^5$He systems inside a harmonic oscillator trap.
We find that the N$^2$LO two-pion exchange 3NF plays a key role in the proper description of the neutron-$\alpha$ $D$-wave phase shifts.
Using the improved values of the $\pi$N couplings to effectively include higher-order corrections to the long-range 3NF, the predicted phase shifts agree well with the empirical values.
Our results constitute an exciting and highly nontrivial test of chiral EFT in the few-nucleon sector and
open the way to experimentally probe the large-distance behavior of the 3NF governed by the chiral symmetry of QCD.

The results of our study can be extended in various directions by, e.g., considering  neutron-$\alpha$ scattering in higher partial waves and exploring other few-nucleon processes like $p\alpha$ and $\alpha\alpha$ scattering.
Such studies would be an indispensable asset in the ongoing efforts toward constructing accurate and precise 3NF from chiral EFT.

\begin{acknowledgments}
This work has been supported in part by the National Natural Science Foundation of China (Grants No. 123B2080, No. 12475117, No. 12141501, No. 11935003), by National Key Laboratory of Neutron Science and Technology NST202401016, by National Key R\&D Program of China (Grants No. 2024YFA1612600, No. 2024YFE0109803), by the High-performance Computing Platform of Peking University, by the European Research Council (ERC AdG NuclearTheory, Grant No. 885150),
by the MKW NRW under the funding code NW21-024-A, by JST ERATO (Grant No. JPMJER2304), by JSPS KAKENHI (Grant No. JP20H05636), and by Start-up Funds of Southeast University (Grant No. 4007022506).
We acknowledge the funding support from the State Key Laboratory of Nuclear Physics and Technology, Peking University (Grant No. NPT2023ZX03).
\end{acknowledgments}


\begin{thebibliography}{66}%
\makeatletter
\providecommand \@ifxundefined [1]{%
 \@ifx{#1\undefined}
}%
\providecommand \@ifnum [1]{%
 \ifnum #1\expandafter \@firstoftwo
 \else \expandafter \@secondoftwo
 \fi
}%
\providecommand \@ifx [1]{%
 \ifx #1\expandafter \@firstoftwo
 \else \expandafter \@secondoftwo
 \fi
}%
\providecommand \natexlab [1]{#1}%
\providecommand \enquote  [1]{``#1''}%
\providecommand \bibnamefont  [1]{#1}%
\providecommand \bibfnamefont [1]{#1}%
\providecommand \citenamefont [1]{#1}%
\providecommand \href@noop [0]{\@secondoftwo}%
\providecommand \href [0]{\begingroup \@sanitize@url \@href}%
\providecommand \@href[1]{\@@startlink{#1}\@@href}%
\providecommand \@@href[1]{\endgroup#1\@@endlink}%
\providecommand \@sanitize@url [0]{\catcode `\\12\catcode `\$12\catcode `\&12\catcode `\#12\catcode `\^12\catcode `\_12\catcode `\%12\relax}%
\providecommand \@@startlink[1]{}%
\providecommand \@@endlink[0]{}%
\providecommand \url  [0]{\begingroup\@sanitize@url \@url }%
\providecommand \@url [1]{\endgroup\@href {#1}{\urlprefix }}%
\providecommand \urlprefix  [0]{URL }%
\providecommand \Eprint [0]{\href }%
\providecommand \doibase [0]{https://doi.org/}%
\providecommand \selectlanguage [0]{\@gobble}%
\providecommand \bibinfo  [0]{\@secondoftwo}%
\providecommand \bibfield  [0]{\@secondoftwo}%
\providecommand \translation [1]{[#1]}%
\providecommand \BibitemOpen [0]{}%
\providecommand \bibitemStop [0]{}%
\providecommand \bibitemNoStop [0]{.\EOS\space}%
\providecommand \EOS [0]{\spacefactor3000\relax}%
\providecommand \BibitemShut  [1]{\csname bibitem#1\endcsname}%
\let\auto@bib@innerbib\@empty
\bibitem [{\citenamefont {Weinberg}(1990)}]{Weinberg1990Phys.Lett.B288}%
  \BibitemOpen
  \bibfield  {author} {\bibinfo {author} {\bibfnamefont {S.}~\bibnamefont {Weinberg}},\ }\bibfield  {title} {\bibinfo {title} {{Nuclear forces from chiral lagrangians}},\ }\href {https://doi.org/https://doi.org/10.1016/0370-2693(90)90938-3} {\bibfield  {journal} {\bibinfo  {journal} {Phys. Lett. B}\ }\textbf {\bibinfo {volume} {251}},\ \bibinfo {pages} {288} (\bibinfo {year} {1990})}\BibitemShut {NoStop}%
\bibitem [{\citenamefont {Weinberg}(1991)}]{Weinberg1991Nucl.Phys.B318}%
  \BibitemOpen
  \bibfield  {author} {\bibinfo {author} {\bibfnamefont {S.}~\bibnamefont {Weinberg}},\ }\bibfield  {title} {\bibinfo {title} {{Effective chiral lagrangians for nucleon-pion interactions and nuclear forces}},\ }\href {https://doi.org/https://doi.org/10.1016/0550-3213(91)90231-L} {\bibfield  {journal} {\bibinfo  {journal} {Nucl. Phys. B}\ }\textbf {\bibinfo {volume} {363}},\ \bibinfo {pages} {3} (\bibinfo {year} {1991})}\BibitemShut {NoStop}%
\bibitem [{\citenamefont {Gross}\ \emph {et~al.}(2023)\citenamefont {Gross} \emph {et~al.}}]{Gross:2022hyw}%
  \BibitemOpen
  \bibfield  {author} {\bibinfo {author} {\bibfnamefont {F.}~\bibnamefont {Gross}} \emph {et~al.},\ }\bibfield  {title} {\bibinfo {title} {{50 Years of Quantum Chromodynamics}},\ }\href {https://doi.org/10.1140/epjc/s10052-023-11949-2} {\bibfield  {journal} {\bibinfo  {journal} {Eur. Phys. J. C}\ }\textbf {\bibinfo {volume} {83}},\ \bibinfo {pages} {1125} (\bibinfo {year} {2023})}\BibitemShut {NoStop}%
\bibitem [{\citenamefont {Epelbaum}\ \emph {et~al.}(2009)\citenamefont {Epelbaum}, \citenamefont {Hammer},\ and\ \citenamefont {Mei\ss{}ner}}]{Epelbaum2009Rev.Mod.Phys.1773}%
  \BibitemOpen
  \bibfield  {author} {\bibinfo {author} {\bibfnamefont {E.}~\bibnamefont {Epelbaum}}, \bibinfo {author} {\bibfnamefont {H.-W.}\ \bibnamefont {Hammer}},\ and\ \bibinfo {author} {\bibfnamefont {U.-G.}\ \bibnamefont {Mei\ss{}ner}},\ }\bibfield  {title} {\bibinfo {title} {{Modern theory of nuclear forces}},\ }\href {https://doi.org/10.1103/RevModPhys.81.1773} {\bibfield  {journal} {\bibinfo  {journal} {Rev. Mod. Phys.}\ }\textbf {\bibinfo {volume} {81}},\ \bibinfo {pages} {1773} (\bibinfo {year} {2009})}\BibitemShut {NoStop}%
\bibitem [{\citenamefont {Machleidt}\ and\ \citenamefont {Entem}(2011)}]{Machleidt2011Phys.Rep.175}%
  \BibitemOpen
  \bibfield  {author} {\bibinfo {author} {\bibfnamefont {R.}~\bibnamefont {Machleidt}}\ and\ \bibinfo {author} {\bibfnamefont {D.}~\bibnamefont {Entem}},\ }\bibfield  {title} {\bibinfo {title} {{Chiral effective field theory and nuclear forces}},\ }\href {https://doi.org/10.1016/j.physrep.2011.02.001} {\bibfield  {journal} {\bibinfo  {journal} {Phys. Rep.}\ }\textbf {\bibinfo {volume} {503}},\ \bibinfo {pages} {1} (\bibinfo {year} {2011})}\BibitemShut {NoStop}%
\bibitem [{\citenamefont {Hammer}\ \emph {et~al.}(2020)\citenamefont {Hammer}, \citenamefont {K\"onig},\ and\ \citenamefont {van Kolck}}]{Hammer2020Rev.Mod.Phys.025004}%
  \BibitemOpen
  \bibfield  {author} {\bibinfo {author} {\bibfnamefont {H.-W.}\ \bibnamefont {Hammer}}, \bibinfo {author} {\bibfnamefont {S.}~\bibnamefont {K\"onig}},\ and\ \bibinfo {author} {\bibfnamefont {U.}~\bibnamefont {van Kolck}},\ }\bibfield  {title} {\bibinfo {title} {{Nuclear effective field theory: Status and perspectives}},\ }\href {https://doi.org/10.1103/RevModPhys.92.025004} {\bibfield  {journal} {\bibinfo  {journal} {Rev. Mod. Phys.}\ }\textbf {\bibinfo {volume} {92}},\ \bibinfo {pages} {025004} (\bibinfo {year} {2020})}\BibitemShut {NoStop}%
\bibitem [{\citenamefont {Epelbaum}(2024)}]{Epelbaum:2024gfg}%
  \BibitemOpen
  \bibfield  {author} {\bibinfo {author} {\bibfnamefont {E.}~\bibnamefont {Epelbaum}},\ }\bibfield  {title} {\bibinfo {title} {{Chiral Symmetry and Nuclear Interactions}},\ }\href {https://doi.org/10.1007/s00601-024-01918-0} {\bibfield  {journal} {\bibinfo  {journal} {Few Body Syst.}\ }\textbf {\bibinfo {volume} {65}},\ \bibinfo {pages} {39} (\bibinfo {year} {2024})}\BibitemShut {NoStop}%
\bibitem [{\citenamefont {Kaiser}\ \emph {et~al.}(1997)\citenamefont {Kaiser}, \citenamefont {Brockmann},\ and\ \citenamefont {Weise}}]{Kaiser1997Nucl.Phys.A758}%
  \BibitemOpen
  \bibfield  {author} {\bibinfo {author} {\bibfnamefont {N.}~\bibnamefont {Kaiser}}, \bibinfo {author} {\bibfnamefont {R.}~\bibnamefont {Brockmann}},\ and\ \bibinfo {author} {\bibfnamefont {W.}~\bibnamefont {Weise}},\ }\bibfield  {title} {\bibinfo {title} {{Peripheral nucleon-nucleon phase shifts and chiral symmetry}},\ }\href {https://doi.org/https://doi.org/10.1016/S0375-9474(97)00586-1} {\bibfield  {journal} {\bibinfo  {journal} {Nucl. Phys. A}\ }\textbf {\bibinfo {volume} {625}},\ \bibinfo {pages} {758} (\bibinfo {year} {1997})}\BibitemShut {NoStop}%
\bibitem [{\citenamefont {Entem}\ and\ \citenamefont {Machleidt}(2002)}]{Entem2002Phys.Rev.C014002}%
  \BibitemOpen
  \bibfield  {author} {\bibinfo {author} {\bibfnamefont {D.~R.}\ \bibnamefont {Entem}}\ and\ \bibinfo {author} {\bibfnamefont {R.}~\bibnamefont {Machleidt}},\ }\bibfield  {title} {\bibinfo {title} {{Chiral $2\ensuremath{\pi}$ exchange at fourth order and peripheral $\mathrm{NN}$ scattering}},\ }\href {https://doi.org/10.1103/PhysRevC.66.014002} {\bibfield  {journal} {\bibinfo  {journal} {Phys. Rev. C}\ }\textbf {\bibinfo {volume} {66}},\ \bibinfo {pages} {014002} (\bibinfo {year} {2002})}\BibitemShut {NoStop}%
\bibitem [{\citenamefont {Birse}\ and\ \citenamefont {McGovern}(2004)}]{Birse:2003nz}%
  \BibitemOpen
  \bibfield  {author} {\bibinfo {author} {\bibfnamefont {M.~C.}\ \bibnamefont {Birse}}\ and\ \bibinfo {author} {\bibfnamefont {J.~A.}\ \bibnamefont {McGovern}},\ }\bibfield  {title} {\bibinfo {title} {{On the effectiveness of effective field theory in peripheral nucleon nucleon scattering}},\ }\href {https://doi.org/10.1103/PhysRevC.70.054002} {\bibfield  {journal} {\bibinfo  {journal} {Phys. Rev. C}\ }\textbf {\bibinfo {volume} {70}},\ \bibinfo {pages} {054002} (\bibinfo {year} {2004})}\BibitemShut {NoStop}%
\bibitem [{\citenamefont {Epelbaum}\ \emph {et~al.}(2015)\citenamefont {Epelbaum}, \citenamefont {Krebs},\ and\ \citenamefont {Mei\ss{}ner}}]{Epelbaum2015Phys.Rev.Lett.122301}%
  \BibitemOpen
  \bibfield  {author} {\bibinfo {author} {\bibfnamefont {E.}~\bibnamefont {Epelbaum}}, \bibinfo {author} {\bibfnamefont {H.}~\bibnamefont {Krebs}},\ and\ \bibinfo {author} {\bibfnamefont {U.-G.}\ \bibnamefont {Mei\ss{}ner}},\ }\bibfield  {title} {\bibinfo {title} {{Precision Nucleon-Nucleon Potential at Fifth Order in the Chiral Expansion}},\ }\href {https://doi.org/10.1103/PhysRevLett.115.122301} {\bibfield  {journal} {\bibinfo  {journal} {Phys. Rev. Lett.}\ }\textbf {\bibinfo {volume} {115}},\ \bibinfo {pages} {122301} (\bibinfo {year} {2015})}\BibitemShut {NoStop}%
\bibitem [{\citenamefont {Reinert}\ \emph {et~al.}(2018)\citenamefont {Reinert}, \citenamefont {Krebs},\ and\ \citenamefont {Epelbaum}}]{Reinert2018Eur.Phys.J.A86}%
  \BibitemOpen
  \bibfield  {author} {\bibinfo {author} {\bibfnamefont {P.}~\bibnamefont {Reinert}}, \bibinfo {author} {\bibfnamefont {H.}~\bibnamefont {Krebs}},\ and\ \bibinfo {author} {\bibfnamefont {E.}~\bibnamefont {Epelbaum}},\ }\bibfield  {title} {\bibinfo {title} {{Semilocal momentum-space regularized chiral two-nucleon potentials up to fifth order}},\ }\href {https://doi.org/10.1140/epja/i2018-12516-4} {\bibfield  {journal} {\bibinfo  {journal} {Eur. Phys. J. A}\ }\textbf {\bibinfo {volume} {54}},\ \bibinfo {pages} {86} (\bibinfo {year} {2018})}\BibitemShut {NoStop}%
\bibitem [{\citenamefont {Kalantar-Nayestanaki}\ \emph {et~al.}(2012)\citenamefont {Kalantar-Nayestanaki}, \citenamefont {Epelbaum}, \citenamefont {Messchendorp},\ and\ \citenamefont {Nogga}}]{Kalantar-Nayestanaki:2011rzs}%
  \BibitemOpen
  \bibfield  {author} {\bibinfo {author} {\bibfnamefont {N.}~\bibnamefont {Kalantar-Nayestanaki}}, \bibinfo {author} {\bibfnamefont {E.}~\bibnamefont {Epelbaum}}, \bibinfo {author} {\bibfnamefont {J.~G.}\ \bibnamefont {Messchendorp}},\ and\ \bibinfo {author} {\bibfnamefont {A.}~\bibnamefont {Nogga}},\ }\bibfield  {title} {\bibinfo {title} {{Signatures of three-nucleon interactions in few-nucleon systems}},\ }\href {https://doi.org/10.1088/0034-4885/75/1/016301} {\bibfield  {journal} {\bibinfo  {journal} {Rept. Prog. Phys.}\ }\textbf {\bibinfo {volume} {75}},\ \bibinfo {pages} {016301} (\bibinfo {year} {2012})}\BibitemShut {NoStop}%
\bibitem [{\citenamefont {Hammer}\ \emph {et~al.}(2013)\citenamefont {Hammer}, \citenamefont {Nogga},\ and\ \citenamefont {Schwenk}}]{Hammer:2012id}%
  \BibitemOpen
  \bibfield  {author} {\bibinfo {author} {\bibfnamefont {H.-W.}\ \bibnamefont {Hammer}}, \bibinfo {author} {\bibfnamefont {A.}~\bibnamefont {Nogga}},\ and\ \bibinfo {author} {\bibfnamefont {A.}~\bibnamefont {Schwenk}},\ }\bibfield  {title} {\bibinfo {title} {{Three-body forces: From cold atoms to nuclei}},\ }\href {https://doi.org/10.1103/RevModPhys.85.197} {\bibfield  {journal} {\bibinfo  {journal} {Rev. Mod. Phys.}\ }\textbf {\bibinfo {volume} {85}},\ \bibinfo {pages} {197} (\bibinfo {year} {2013})}\BibitemShut {NoStop}%
\bibitem [{\citenamefont {Endo}\ \emph {et~al.}(2025)\citenamefont {Endo}, \citenamefont {Epelbaum}, \citenamefont {Naidon}, \citenamefont {Nishida}, \citenamefont {Sekiguchi},\ and\ \citenamefont {Takahashi}}]{Endo:2024cbz}%
  \BibitemOpen
  \bibfield  {author} {\bibinfo {author} {\bibfnamefont {S.}~\bibnamefont {Endo}}, \bibinfo {author} {\bibfnamefont {E.}~\bibnamefont {Epelbaum}}, \bibinfo {author} {\bibfnamefont {P.}~\bibnamefont {Naidon}}, \bibinfo {author} {\bibfnamefont {Y.}~\bibnamefont {Nishida}}, \bibinfo {author} {\bibfnamefont {K.}~\bibnamefont {Sekiguchi}},\ and\ \bibinfo {author} {\bibfnamefont {Y.}~\bibnamefont {Takahashi}},\ }\bibfield  {title} {\bibinfo {title} {{Three-body forces and Efimov physics in nuclei and atoms}},\ }\href {https://doi.org/10.1140/epja/s10050-024-01467-4} {\bibfield  {journal} {\bibinfo  {journal} {Eur. Phys. J. A}\ }\textbf {\bibinfo {volume} {61}},\ \bibinfo {pages} {9} (\bibinfo {year} {2025})}\BibitemShut {NoStop}%
\bibitem [{\citenamefont {van Kolck}(1994)}]{vanKolck1994Phys.Rev.C2932}%
  \BibitemOpen
  \bibfield  {author} {\bibinfo {author} {\bibfnamefont {U.}~\bibnamefont {van Kolck}},\ }\bibfield  {title} {\bibinfo {title} {{Few-nucleon forces from chiral Lagrangians}},\ }\href {https://doi.org/10.1103/PhysRevC.49.2932} {\bibfield  {journal} {\bibinfo  {journal} {Phys. Rev. C}\ }\textbf {\bibinfo {volume} {49}},\ \bibinfo {pages} {2932} (\bibinfo {year} {1994})}\BibitemShut {NoStop}%
\bibitem [{\citenamefont {Epelbaum}\ \emph {et~al.}(2002)\citenamefont {Epelbaum}, \citenamefont {Nogga}, \citenamefont {Gl\"ockle}, \citenamefont {Kamada}, \citenamefont {Mei\ss{}ner},\ and\ \citenamefont {Wita\l{}a}}]{Epelbaum2002Phys.Rev.C064001}%
  \BibitemOpen
  \bibfield  {author} {\bibinfo {author} {\bibfnamefont {E.}~\bibnamefont {Epelbaum}}, \bibinfo {author} {\bibfnamefont {A.}~\bibnamefont {Nogga}}, \bibinfo {author} {\bibfnamefont {W.}~\bibnamefont {Gl\"ockle}}, \bibinfo {author} {\bibfnamefont {H.}~\bibnamefont {Kamada}}, \bibinfo {author} {\bibfnamefont {U.-G.}\ \bibnamefont {Mei\ss{}ner}},\ and\ \bibinfo {author} {\bibfnamefont {H.}~\bibnamefont {Wita\l{}a}},\ }\bibfield  {title} {\bibinfo {title} {{Three-nucleon forces from chiral effective field theory}},\ }\href {https://doi.org/10.1103/PhysRevC.66.064001} {\bibfield  {journal} {\bibinfo  {journal} {Phys. Rev. C}\ }\textbf {\bibinfo {volume} {66}},\ \bibinfo {pages} {064001} (\bibinfo {year} {2002})}\BibitemShut {NoStop}%
\bibitem [{\citenamefont {Krebs}\ and\ \citenamefont {Epelbaum}(2024{\natexlab{a}})}]{Krebs:2023ljo}%
  \BibitemOpen
  \bibfield  {author} {\bibinfo {author} {\bibfnamefont {H.}~\bibnamefont {Krebs}}\ and\ \bibinfo {author} {\bibfnamefont {E.}~\bibnamefont {Epelbaum}},\ }\bibfield  {title} {\bibinfo {title} {{Toward consistent nuclear interactions from chiral Lagrangians. I. The path-integral approach}},\ }\href {https://doi.org/10.1103/PhysRevC.110.044003} {\bibfield  {journal} {\bibinfo  {journal} {Phys. Rev. C}\ }\textbf {\bibinfo {volume} {110}},\ \bibinfo {pages} {044003} (\bibinfo {year} {2024}{\natexlab{a}})}\BibitemShut {NoStop}%
\bibitem [{\citenamefont {Krebs}\ and\ \citenamefont {Epelbaum}(2024{\natexlab{b}})}]{Krebs:2023gge}%
  \BibitemOpen
  \bibfield  {author} {\bibinfo {author} {\bibfnamefont {H.}~\bibnamefont {Krebs}}\ and\ \bibinfo {author} {\bibfnamefont {E.}~\bibnamefont {Epelbaum}},\ }\bibfield  {title} {\bibinfo {title} {{Toward consistent nuclear interactions from chiral Lagrangians. II. Symmetry preserving regularization}},\ }\href {https://doi.org/10.1103/PhysRevC.110.044004} {\bibfield  {journal} {\bibinfo  {journal} {Phys. Rev. C}\ }\textbf {\bibinfo {volume} {110}},\ \bibinfo {pages} {044004} (\bibinfo {year} {2024}{\natexlab{b}})}\BibitemShut {NoStop}%
\bibitem [{\citenamefont {Barreiro}\ \emph {et~al.}(1998)\citenamefont {Barreiro}, \citenamefont {Higa}, \citenamefont {Lima},\ and\ \citenamefont {Robilotta}}]{Barreiro1998Phys.Rev.C21422148}%
  \BibitemOpen
  \bibfield  {author} {\bibinfo {author} {\bibfnamefont {L.~A.}\ \bibnamefont {Barreiro}}, \bibinfo {author} {\bibfnamefont {R.}~\bibnamefont {Higa}}, \bibinfo {author} {\bibfnamefont {C.~L.}\ \bibnamefont {Lima}},\ and\ \bibinfo {author} {\bibfnamefont {M.~R.}\ \bibnamefont {Robilotta}},\ }\bibfield  {title} {\bibinfo {title} {{Peripheral $N\ensuremath{\alpha}$ scattering: A tool for identifying the two pion exchange component of the $\mathrm{NN}$ potential}},\ }\href {https://doi.org/10.1103/PhysRevC.57.2142} {\bibfield  {journal} {\bibinfo  {journal} {Phys. Rev. C}\ }\textbf {\bibinfo {volume} {57}},\ \bibinfo {pages} {2142} (\bibinfo {year} {1998})}\BibitemShut {NoStop}%
\bibitem [{\citenamefont {Higa}\ and\ \citenamefont {Robilotta}(1999)}]{Higa1999arXiv9908062}%
  \BibitemOpen
  \bibfield  {author} {\bibinfo {author} {\bibfnamefont {R.}~\bibnamefont {Higa}}\ and\ \bibinfo {author} {\bibfnamefont {M.~R.}\ \bibnamefont {Robilotta}},\ }\bibfield  {title} {\bibinfo {title} {{Three body force in peripheral $N\alpha$ scattering}},\ }\href@noop {} {\bibfield  {journal} {\bibinfo  {journal} {arXiv:nucl-th/9908062}\ } (\bibinfo {year} {1999})}\BibitemShut {NoStop}%
\bibitem [{\citenamefont {Carlson}\ \emph {et~al.}(1987)\citenamefont {Carlson}, \citenamefont {Schmidt},\ and\ \citenamefont {Kalos}}]{Carlson1987Phys.Rev.C2731}%
  \BibitemOpen
  \bibfield  {author} {\bibinfo {author} {\bibfnamefont {J.}~\bibnamefont {Carlson}}, \bibinfo {author} {\bibfnamefont {K.~E.}\ \bibnamefont {Schmidt}},\ and\ \bibinfo {author} {\bibfnamefont {M.~H.}\ \bibnamefont {Kalos}},\ }\bibfield  {title} {\bibinfo {title} {{Microscopic calculations of $^{5}\mathrm{He}$ with realistic interactions}},\ }\href {https://doi.org/10.1103/PhysRevC.36.27} {\bibfield  {journal} {\bibinfo  {journal} {Phys. Rev. C}\ }\textbf {\bibinfo {volume} {36}},\ \bibinfo {pages} {27} (\bibinfo {year} {1987})}\BibitemShut {NoStop}%
\bibitem [{\citenamefont {Nollett}\ \emph {et~al.}(2007)\citenamefont {Nollett}, \citenamefont {Pieper}, \citenamefont {Wiringa}, \citenamefont {Carlson},\ and\ \citenamefont {Hale}}]{Nollett2007Phys.Rev.Lett.022502}%
  \BibitemOpen
  \bibfield  {author} {\bibinfo {author} {\bibfnamefont {K.~M.}\ \bibnamefont {Nollett}}, \bibinfo {author} {\bibfnamefont {S.~C.}\ \bibnamefont {Pieper}}, \bibinfo {author} {\bibfnamefont {R.~B.}\ \bibnamefont {Wiringa}}, \bibinfo {author} {\bibfnamefont {J.}~\bibnamefont {Carlson}},\ and\ \bibinfo {author} {\bibfnamefont {G.~M.}\ \bibnamefont {Hale}},\ }\bibfield  {title} {\bibinfo {title} {{Quantum Monte Carlo Calculations of Neutron-$\ensuremath{\alpha}$ Scattering}},\ }\href {https://doi.org/10.1103/PhysRevLett.99.022502} {\bibfield  {journal} {\bibinfo  {journal} {Phys. Rev. Lett.}\ }\textbf {\bibinfo {volume} {99}},\ \bibinfo {pages} {022502} (\bibinfo {year} {2007})}\BibitemShut {NoStop}%
\bibitem [{\citenamefont {Lynn}\ \emph {et~al.}(2016)\citenamefont {Lynn}, \citenamefont {Tews}, \citenamefont {Carlson}, \citenamefont {Gandolfi}, \citenamefont {Gezerlis}, \citenamefont {Schmidt},\ and\ \citenamefont {Schwenk}}]{Lynn2016Phys.Rev.Lett.062501}%
  \BibitemOpen
  \bibfield  {author} {\bibinfo {author} {\bibfnamefont {J.~E.}\ \bibnamefont {Lynn}}, \bibinfo {author} {\bibfnamefont {I.}~\bibnamefont {Tews}}, \bibinfo {author} {\bibfnamefont {J.}~\bibnamefont {Carlson}}, \bibinfo {author} {\bibfnamefont {S.}~\bibnamefont {Gandolfi}}, \bibinfo {author} {\bibfnamefont {A.}~\bibnamefont {Gezerlis}}, \bibinfo {author} {\bibfnamefont {K.~E.}\ \bibnamefont {Schmidt}},\ and\ \bibinfo {author} {\bibfnamefont {A.}~\bibnamefont {Schwenk}},\ }\bibfield  {title} {\bibinfo {title} {{Chiral Three-Nucleon Interactions in Light Nuclei, Neutron-$\ensuremath{\alpha}$ Scattering, and Neutron Matter}},\ }\href {https://doi.org/10.1103/PhysRevLett.116.062501} {\bibfield  {journal} {\bibinfo  {journal} {Phys. Rev. Lett.}\ }\textbf {\bibinfo {volume} {116}},\ \bibinfo {pages} {062501} (\bibinfo {year} {2016})}\BibitemShut {NoStop}%
\bibitem [{\citenamefont {Flores}\ \emph {et~al.}(2025)\citenamefont {Flores}, \citenamefont {Nollett},\ and\ \citenamefont {Piarulli}}]{Flores2025PRC112.014008}%
  \BibitemOpen
  \bibfield  {author} {\bibinfo {author} {\bibfnamefont {A.~R.}\ \bibnamefont {Flores}}, \bibinfo {author} {\bibfnamefont {K.~M.}\ \bibnamefont {Nollett}},\ and\ \bibinfo {author} {\bibfnamefont {M.}~\bibnamefont {Piarulli}},\ }\bibfield  {title} {\bibinfo {title} {Quantum monte carlo calculations of neutron-$\ensuremath{\alpha}$ scattering via an integral relation},\ }\href {https://doi.org/10.1103/q4dy-vhv1} {\bibfield  {journal} {\bibinfo  {journal} {Phys. Rev. C}\ }\textbf {\bibinfo {volume} {112}},\ \bibinfo {pages} {014008} (\bibinfo {year} {2025})}\BibitemShut {NoStop}%
\bibitem [{\citenamefont {Lazauskas}(2018)}]{Lazauskas2018Phys.Rev.C044002}%
  \BibitemOpen
  \bibfield  {author} {\bibinfo {author} {\bibfnamefont {R.}~\bibnamefont {Lazauskas}},\ }\bibfield  {title} {\bibinfo {title} {Solution of the $n\ensuremath{-}^{4}\mathrm{He}$ elastic scattering problem using the faddeev-yakubovsky equations},\ }\href {https://doi.org/10.1103/PhysRevC.97.044002} {\bibfield  {journal} {\bibinfo  {journal} {Phys. Rev. C}\ }\textbf {\bibinfo {volume} {97}},\ \bibinfo {pages} {044002} (\bibinfo {year} {2018})}\BibitemShut {NoStop}%
\bibitem [{\citenamefont {Lazauskas}\ and\ \citenamefont {Carbonell}(2020)}]{Lazauskas2020Front.Phys.}%
  \BibitemOpen
  \bibfield  {author} {\bibinfo {author} {\bibfnamefont {R.}~\bibnamefont {Lazauskas}}\ and\ \bibinfo {author} {\bibfnamefont {J.}~\bibnamefont {Carbonell}},\ }\bibfield  {title} {\bibinfo {title} {{Description of Four- and Five-Nucleon Systems by Solving Faddeev-Yakubovsky Equations in Configuration Space}},\ }\bibfield  {journal} {\bibinfo  {journal} {Front. Phys.}\ }\textbf {\bibinfo {volume} {7}},\ \href {https://doi.org/10.3389/fphy.2019.00251} {10.3389/fphy.2019.00251} (\bibinfo {year} {2020})\BibitemShut {NoStop}%
\bibitem [{\citenamefont {Kravvaris}\ \emph {et~al.}(2020)\citenamefont {Kravvaris}, \citenamefont {Quinlan}, \citenamefont {Quaglioni}, \citenamefont {Wendt},\ and\ \citenamefont {Navr\'atil}}]{Kravvaris2020Phys.Rev.C024616}%
  \BibitemOpen
  \bibfield  {author} {\bibinfo {author} {\bibfnamefont {K.}~\bibnamefont {Kravvaris}}, \bibinfo {author} {\bibfnamefont {K.~R.}\ \bibnamefont {Quinlan}}, \bibinfo {author} {\bibfnamefont {S.}~\bibnamefont {Quaglioni}}, \bibinfo {author} {\bibfnamefont {K.~A.}\ \bibnamefont {Wendt}},\ and\ \bibinfo {author} {\bibfnamefont {P.}~\bibnamefont {Navr\'atil}},\ }\bibfield  {title} {\bibinfo {title} {{Quantifying uncertainties in neutron-$\ensuremath{\alpha}$ scattering with chiral nucleon-nucleon and three-nucleon forces}},\ }\href {https://doi.org/10.1103/PhysRevC.102.024616} {\bibfield  {journal} {\bibinfo  {journal} {Phys. Rev. C}\ }\textbf {\bibinfo {volume} {102}},\ \bibinfo {pages} {024616} (\bibinfo {year} {2020})}\BibitemShut {NoStop}%
\bibitem [{\citenamefont {Shirokov}\ \emph {et~al.}(2018)\citenamefont {Shirokov}, \citenamefont {Mazur}, \citenamefont {Mazur}, \citenamefont {Mazur}, \citenamefont {Shin}, \citenamefont {Kim}, \citenamefont {Blokhintsev},\ and\ \citenamefont {Vary}}]{Shirokov2018Phys.Rev.C044624}%
  \BibitemOpen
  \bibfield  {author} {\bibinfo {author} {\bibfnamefont {A.~M.}\ \bibnamefont {Shirokov}}, \bibinfo {author} {\bibfnamefont {A.~I.}\ \bibnamefont {Mazur}}, \bibinfo {author} {\bibfnamefont {I.~A.}\ \bibnamefont {Mazur}}, \bibinfo {author} {\bibfnamefont {E.~A.}\ \bibnamefont {Mazur}}, \bibinfo {author} {\bibfnamefont {I.~J.}\ \bibnamefont {Shin}}, \bibinfo {author} {\bibfnamefont {Y.}~\bibnamefont {Kim}}, \bibinfo {author} {\bibfnamefont {L.~D.}\ \bibnamefont {Blokhintsev}},\ and\ \bibinfo {author} {\bibfnamefont {J.~P.}\ \bibnamefont {Vary}},\ }\bibfield  {title} {\bibinfo {title} {{Nucleon-$\ensuremath{\alpha}$ scattering and resonances in $^{5}\mathrm{He}$ and $^{5}\mathrm{Li}$ with JISP16 and Daejeon16 $\mathit{NN}$ interactions}},\ }\href {https://doi.org/10.1103/PhysRevC.98.044624} {\bibfield  {journal} {\bibinfo  {journal} {Phys. Rev. C}\ }\textbf {\bibinfo {volume} {98}},\ \bibinfo {pages} {044624} (\bibinfo {year} {2018})}\BibitemShut {NoStop}%
\bibitem [{\citenamefont {Navr\'atil}\ \emph {et~al.}(2016)\citenamefont {Navr\'atil}, \citenamefont {Quaglioni}, \citenamefont {Hupin}, \citenamefont {Romero-Redondo},\ and\ \citenamefont {Calci}}]{Navratil2016Phys.Scr.053002}%
  \BibitemOpen
  \bibfield  {author} {\bibinfo {author} {\bibfnamefont {P.}~\bibnamefont {Navr\'atil}}, \bibinfo {author} {\bibfnamefont {S.}~\bibnamefont {Quaglioni}}, \bibinfo {author} {\bibfnamefont {G.}~\bibnamefont {Hupin}}, \bibinfo {author} {\bibfnamefont {C.}~\bibnamefont {Romero-Redondo}},\ and\ \bibinfo {author} {\bibfnamefont {A.}~\bibnamefont {Calci}},\ }\bibfield  {title} {\bibinfo {title} {Unified ab initio approaches to nuclear structure and reactions},\ }\href {https://doi.org/10.1088/0031-8949/91/5/053002} {\bibfield  {journal} {\bibinfo  {journal} {Phys. Scr.}\ }\textbf {\bibinfo {volume} {91}},\ \bibinfo {pages} {053002} (\bibinfo {year} {2016})}\BibitemShut {NoStop}%
\bibitem [{\citenamefont {Zhang}\ \emph {et~al.}(2020)\citenamefont {Zhang}, \citenamefont {Stroberg}, \citenamefont {Navr\'atil}, \citenamefont {Gwak}, \citenamefont {Melendez}, \citenamefont {Furnstahl},\ and\ \citenamefont {Holt}}]{Zhang2020Phys.Rev.Lett.112503}%
  \BibitemOpen
  \bibfield  {author} {\bibinfo {author} {\bibfnamefont {X.}~\bibnamefont {Zhang}}, \bibinfo {author} {\bibfnamefont {S.~R.}\ \bibnamefont {Stroberg}}, \bibinfo {author} {\bibfnamefont {P.}~\bibnamefont {Navr\'atil}}, \bibinfo {author} {\bibfnamefont {C.}~\bibnamefont {Gwak}}, \bibinfo {author} {\bibfnamefont {J.~A.}\ \bibnamefont {Melendez}}, \bibinfo {author} {\bibfnamefont {R.~J.}\ \bibnamefont {Furnstahl}},\ and\ \bibinfo {author} {\bibfnamefont {J.~D.}\ \bibnamefont {Holt}},\ }\bibfield  {title} {\bibinfo {title} {{Ab Initio Calculations of Low-Energy Nuclear Scattering Using Confining Potential Traps}},\ }\href {https://doi.org/10.1103/PhysRevLett.125.112503} {\bibfield  {journal} {\bibinfo  {journal} {Phys. Rev. Lett.}\ }\textbf {\bibinfo {volume} {125}},\ \bibinfo {pages} {112503} (\bibinfo {year} {2020})}\BibitemShut {NoStop}%
\bibitem [{\citenamefont {Mercenne}\ \emph {et~al.}(2022)\citenamefont {Mercenne}, \citenamefont {Launey}, \citenamefont {Dytrych}, \citenamefont {Escher}, \citenamefont {Quaglioni}, \citenamefont {Sargsyan}, \citenamefont {Langr},\ and\ \citenamefont {Draayer}}]{Mercenne2022Comput.Phys.Commun.108476}%
  \BibitemOpen
  \bibfield  {author} {\bibinfo {author} {\bibfnamefont {A.}~\bibnamefont {Mercenne}}, \bibinfo {author} {\bibfnamefont {K.}~\bibnamefont {Launey}}, \bibinfo {author} {\bibfnamefont {T.}~\bibnamefont {Dytrych}}, \bibinfo {author} {\bibfnamefont {J.}~\bibnamefont {Escher}}, \bibinfo {author} {\bibfnamefont {S.}~\bibnamefont {Quaglioni}}, \bibinfo {author} {\bibfnamefont {G.}~\bibnamefont {Sargsyan}}, \bibinfo {author} {\bibfnamefont {D.}~\bibnamefont {Langr}},\ and\ \bibinfo {author} {\bibfnamefont {J.}~\bibnamefont {Draayer}},\ }\bibfield  {title} {\bibinfo {title} {{Efficacy of the symmetry-adapted basis for ab initio nucleon-nucleus interactions for light- and intermediate-mass nuclei}},\ }\href {https://doi.org/https://doi.org/10.1016/j.cpc.2022.108476} {\bibfield  {journal} {\bibinfo  {journal} {Comput. Phys. Commun.}\ }\textbf {\bibinfo {volume} {280}},\ \bibinfo {pages} {108476} (\bibinfo {year} {2022})}\BibitemShut {NoStop}%
\bibitem [{\citenamefont {Bagnarol}\ \emph {et~al.}(2023)\citenamefont {Bagnarol}, \citenamefont {Sch\"afer}, \citenamefont {Bazak},\ and\ \citenamefont {Barnea}}]{Bagnarol2023Phys.Lett.B138078}%
  \BibitemOpen
  \bibfield  {author} {\bibinfo {author} {\bibfnamefont {M.}~\bibnamefont {Bagnarol}}, \bibinfo {author} {\bibfnamefont {M.}~\bibnamefont {Sch\"afer}}, \bibinfo {author} {\bibfnamefont {B.}~\bibnamefont {Bazak}},\ and\ \bibinfo {author} {\bibfnamefont {N.}~\bibnamefont {Barnea}},\ }\bibfield  {title} {\bibinfo {title} {Five-body calculation of s-wave n-4he scattering at next-to-leading order pionless effective field theory},\ }\href {https://doi.org/https://doi.org/10.1016/j.physletb.2023.138078} {\bibfield  {journal} {\bibinfo  {journal} {Phys. Lett. B}\ }\textbf {\bibinfo {volume} {844}},\ \bibinfo {pages} {138078} (\bibinfo {year} {2023})}\BibitemShut {NoStop}%
\bibitem [{\citenamefont {Elhatisari}\ \emph {et~al.}(2025)\citenamefont {Elhatisari}, \citenamefont {Hildenbrand},\ and\ \citenamefont {Mei\ss{}ner}}]{elhatisari2025arXiv}%
  \BibitemOpen
  \bibfield  {author} {\bibinfo {author} {\bibfnamefont {S.}~\bibnamefont {Elhatisari}}, \bibinfo {author} {\bibfnamefont {F.}~\bibnamefont {Hildenbrand}},\ and\ \bibinfo {author} {\bibfnamefont {U.-G.}\ \bibnamefont {Mei\ss{}ner}},\ }\href {https://arxiv.org/abs/2507.08495} {\bibinfo {title} {Ab initio lattice study of neutron-alpha scattering with chiral forces at n3lo}} (\bibinfo {year} {2025}),\ \Eprint {https://arxiv.org/abs/2507.08495} {arXiv:2507.08495 [nucl-th]} \BibitemShut {NoStop}%
\bibitem [{\citenamefont {Gezerlis}\ \emph {et~al.}(2013)\citenamefont {Gezerlis}, \citenamefont {Tews}, \citenamefont {Epelbaum}, \citenamefont {Gandolfi}, \citenamefont {Hebeler}, \citenamefont {Nogga},\ and\ \citenamefont {Schwenk}}]{Gezerlis2013Phys.Rev.Lett.032501}%
  \BibitemOpen
  \bibfield  {author} {\bibinfo {author} {\bibfnamefont {A.}~\bibnamefont {Gezerlis}}, \bibinfo {author} {\bibfnamefont {I.}~\bibnamefont {Tews}}, \bibinfo {author} {\bibfnamefont {E.}~\bibnamefont {Epelbaum}}, \bibinfo {author} {\bibfnamefont {S.}~\bibnamefont {Gandolfi}}, \bibinfo {author} {\bibfnamefont {K.}~\bibnamefont {Hebeler}}, \bibinfo {author} {\bibfnamefont {A.}~\bibnamefont {Nogga}},\ and\ \bibinfo {author} {\bibfnamefont {A.}~\bibnamefont {Schwenk}},\ }\bibfield  {title} {\bibinfo {title} {{Quantum Monte Carlo Calculations with Chiral Effective Field Theory Interactions}},\ }\href {https://doi.org/10.1103/PhysRevLett.111.032501} {\bibfield  {journal} {\bibinfo  {journal} {Phys. Rev. Lett.}\ }\textbf {\bibinfo {volume} {111}},\ \bibinfo {pages} {032501} (\bibinfo {year} {2013})}\BibitemShut {NoStop}%
\bibitem [{\citenamefont {Gezerlis}\ \emph {et~al.}(2014)\citenamefont {Gezerlis}, \citenamefont {Tews}, \citenamefont {Epelbaum}, \citenamefont {Freunek}, \citenamefont {Gandolfi}, \citenamefont {Hebeler}, \citenamefont {Nogga},\ and\ \citenamefont {Schwenk}}]{Gezerlis2014Phys.Rev.C054323}%
  \BibitemOpen
  \bibfield  {author} {\bibinfo {author} {\bibfnamefont {A.}~\bibnamefont {Gezerlis}}, \bibinfo {author} {\bibfnamefont {I.}~\bibnamefont {Tews}}, \bibinfo {author} {\bibfnamefont {E.}~\bibnamefont {Epelbaum}}, \bibinfo {author} {\bibfnamefont {M.}~\bibnamefont {Freunek}}, \bibinfo {author} {\bibfnamefont {S.}~\bibnamefont {Gandolfi}}, \bibinfo {author} {\bibfnamefont {K.}~\bibnamefont {Hebeler}}, \bibinfo {author} {\bibfnamefont {A.}~\bibnamefont {Nogga}},\ and\ \bibinfo {author} {\bibfnamefont {A.}~\bibnamefont {Schwenk}},\ }\bibfield  {title} {\bibinfo {title} {{Local chiral effective field theory interactions and quantum Monte Carlo applications}},\ }\href {https://doi.org/10.1103/PhysRevC.90.054323} {\bibfield  {journal} {\bibinfo  {journal} {Phys. Rev. C}\ }\textbf {\bibinfo {volume} {90}},\ \bibinfo {pages} {054323} (\bibinfo {year} {2014})}\BibitemShut {NoStop}%
\bibitem [{\citenamefont {Tews}\ \emph {et~al.}(2016)\citenamefont {Tews}, \citenamefont {Gandolfi}, \citenamefont {Gezerlis},\ and\ \citenamefont {Schwenk}}]{Tews2016Phys.Rev.C024305}%
  \BibitemOpen
  \bibfield  {author} {\bibinfo {author} {\bibfnamefont {I.}~\bibnamefont {Tews}}, \bibinfo {author} {\bibfnamefont {S.}~\bibnamefont {Gandolfi}}, \bibinfo {author} {\bibfnamefont {A.}~\bibnamefont {Gezerlis}},\ and\ \bibinfo {author} {\bibfnamefont {A.}~\bibnamefont {Schwenk}},\ }\bibfield  {title} {\bibinfo {title} {{Quantum Monte Carlo calculations of neutron matter with chiral three-body forces}},\ }\href {https://doi.org/10.1103/PhysRevC.93.024305} {\bibfield  {journal} {\bibinfo  {journal} {Phys. Rev. C}\ }\textbf {\bibinfo {volume} {93}},\ \bibinfo {pages} {024305} (\bibinfo {year} {2016})}\BibitemShut {NoStop}%
\bibitem [{\citenamefont {Lynn}\ \emph {et~al.}(2017)\citenamefont {Lynn}, \citenamefont {Tews}, \citenamefont {Carlson}, \citenamefont {Gandolfi}, \citenamefont {Gezerlis}, \citenamefont {Schmidt},\ and\ \citenamefont {Schwenk}}]{Lynn2017Phys.Rev.C054007}%
  \BibitemOpen
  \bibfield  {author} {\bibinfo {author} {\bibfnamefont {J.~E.}\ \bibnamefont {Lynn}}, \bibinfo {author} {\bibfnamefont {I.}~\bibnamefont {Tews}}, \bibinfo {author} {\bibfnamefont {J.}~\bibnamefont {Carlson}}, \bibinfo {author} {\bibfnamefont {S.}~\bibnamefont {Gandolfi}}, \bibinfo {author} {\bibfnamefont {A.}~\bibnamefont {Gezerlis}}, \bibinfo {author} {\bibfnamefont {K.~E.}\ \bibnamefont {Schmidt}},\ and\ \bibinfo {author} {\bibfnamefont {A.}~\bibnamefont {Schwenk}},\ }\bibfield  {title} {\bibinfo {title} {{Quantum Monte Carlo calculations of light nuclei with local chiral two- and three-nucleon interactions}},\ }\href {https://doi.org/10.1103/PhysRevC.96.054007} {\bibfield  {journal} {\bibinfo  {journal} {Phys. Rev. C}\ }\textbf {\bibinfo {volume} {96}},\ \bibinfo {pages} {054007} (\bibinfo {year} {2017})}\BibitemShut {NoStop}%
\bibitem [{\citenamefont {Epelbaum}\ \emph {et~al.}(2005)\citenamefont {Epelbaum}, \citenamefont {Gl\"ockle},\ and\ \citenamefont {Mei\ss{}ner}}]{Epelbaum2005Nucl.Phys.A362}%
  \BibitemOpen
  \bibfield  {author} {\bibinfo {author} {\bibfnamefont {E.}~\bibnamefont {Epelbaum}}, \bibinfo {author} {\bibfnamefont {W.}~\bibnamefont {Gl\"ockle}},\ and\ \bibinfo {author} {\bibfnamefont {U.-G.}\ \bibnamefont {Mei\ss{}ner}},\ }\bibfield  {title} {\bibinfo {title} {{The two-nucleon system at next-to-next-to-next-to-leading order}},\ }\href {https://doi.org/https://doi.org/10.1016/j.nuclphysa.2004.09.107} {\bibfield  {journal} {\bibinfo  {journal} {Nucl. Phys. A}\ }\textbf {\bibinfo {volume} {747}},\ \bibinfo {pages} {362} (\bibinfo {year} {2005})}\BibitemShut {NoStop}%
\bibitem [{\citenamefont {B\"uttiker}\ and\ \citenamefont {Mei\ss{}ner}(2000)}]{Buettiker2000Nucl.Phys.A97}%
  \BibitemOpen
  \bibfield  {author} {\bibinfo {author} {\bibfnamefont {P.}~\bibnamefont {B\"uttiker}}\ and\ \bibinfo {author} {\bibfnamefont {U.-G.}\ \bibnamefont {Mei\ss{}ner}},\ }\bibfield  {title} {\bibinfo {title} {{Pion–nucleon scattering inside the Mandelstam triangle}},\ }\href {https://doi.org/https://doi.org/10.1016/S0375-9474(99)00813-1} {\bibfield  {journal} {\bibinfo  {journal} {Nucl. Phys. A}\ }\textbf {\bibinfo {volume} {668}},\ \bibinfo {pages} {97} (\bibinfo {year} {2000})}\BibitemShut {NoStop}%
\bibitem [{\citenamefont {Busch}\ \emph {et~al.}(1998)\citenamefont {Busch}, \citenamefont {Englert}, \citenamefont {Rza\.{z}ewski},\ and\ \citenamefont {Wilkens}}]{Busch1998FoundationsofPhysics549}%
  \BibitemOpen
  \bibfield  {author} {\bibinfo {author} {\bibfnamefont {T.}~\bibnamefont {Busch}}, \bibinfo {author} {\bibfnamefont {B.-G.}\ \bibnamefont {Englert}}, \bibinfo {author} {\bibfnamefont {K.}~\bibnamefont {Rza\.{z}ewski}},\ and\ \bibinfo {author} {\bibfnamefont {M.}~\bibnamefont {Wilkens}},\ }\bibfield  {title} {\bibinfo {title} {{Two Cold Atoms in a Harmonic Trap}},\ }\href {https://doi.org/10.1023/A:1018705520999} {\bibfield  {journal} {\bibinfo  {journal} {Foundations of Physics}\ }\textbf {\bibinfo {volume} {28}},\ \bibinfo {pages} {549} (\bibinfo {year} {1998})}\BibitemShut {NoStop}%
\bibitem [{\citenamefont {Suzuki}\ \emph {et~al.}(2009)\citenamefont {Suzuki}, \citenamefont {Liang},\ and\ \citenamefont {Bhaduri}}]{Suzuki2009Phys.Rev.A033601}%
  \BibitemOpen
  \bibfield  {author} {\bibinfo {author} {\bibfnamefont {A.}~\bibnamefont {Suzuki}}, \bibinfo {author} {\bibfnamefont {Y.}~\bibnamefont {Liang}},\ and\ \bibinfo {author} {\bibfnamefont {R.~K.}\ \bibnamefont {Bhaduri}},\ }\bibfield  {title} {\bibinfo {title} {{Two-atom energy spectrum in a harmonic trap near a Feshbach resonance at higher partial waves}},\ }\href {https://doi.org/10.1103/PhysRevA.80.033601} {\bibfield  {journal} {\bibinfo  {journal} {Phys. Rev. A}\ }\textbf {\bibinfo {volume} {80}},\ \bibinfo {pages} {033601} (\bibinfo {year} {2009})}\BibitemShut {NoStop}%
\bibitem [{\citenamefont {Massella}\ \emph {et~al.}(2020)\citenamefont {Massella}, \citenamefont {Barranco}, \citenamefont {Lonardoni}, \citenamefont {Lovato}, \citenamefont {Pederiva},\ and\ \citenamefont {Vigezzi}}]{Massella2020J.Phys.G47.035105}%
  \BibitemOpen
  \bibfield  {author} {\bibinfo {author} {\bibfnamefont {P.}~\bibnamefont {Massella}}, \bibinfo {author} {\bibfnamefont {F.}~\bibnamefont {Barranco}}, \bibinfo {author} {\bibfnamefont {D.}~\bibnamefont {Lonardoni}}, \bibinfo {author} {\bibfnamefont {A.}~\bibnamefont {Lovato}}, \bibinfo {author} {\bibfnamefont {F.}~\bibnamefont {Pederiva}},\ and\ \bibinfo {author} {\bibfnamefont {E.}~\bibnamefont {Vigezzi}},\ }\bibfield  {title} {\bibinfo {title} {Exact restoration of galilei invariance in density functional calculations with quantum monte carlo},\ }\href {https://doi.org/10.1088/1361-6471/ab588c} {\bibfield  {journal} {\bibinfo  {journal} {J. Phys. G}\ }\textbf {\bibinfo {volume} {47}},\ \bibinfo {pages} {035105} (\bibinfo {year} {2020})}\BibitemShut {NoStop}%
\bibitem [{\citenamefont {Bond}\ and\ \citenamefont {Firk}(1977)}]{Bond1977Nucl.Phys.A317}%
  \BibitemOpen
  \bibfield  {author} {\bibinfo {author} {\bibfnamefont {J.}~\bibnamefont {Bond}}\ and\ \bibinfo {author} {\bibfnamefont {F.}~\bibnamefont {Firk}},\ }\bibfield  {title} {\bibinfo {title} {{Determination of R-function and physical-state parameters for n-4He elastic scattering below 21 MeV}},\ }\href {https://doi.org/https://doi.org/10.1016/0375-9474(77)90499-7} {\bibfield  {journal} {\bibinfo  {journal} {Nucl. Phys. A}\ }\textbf {\bibinfo {volume} {287}},\ \bibinfo {pages} {317} (\bibinfo {year} {1977})}\BibitemShut {NoStop}%
\bibitem [{\citenamefont {Adams}\ \emph {et~al.}(2021)\citenamefont {Adams}, \citenamefont {Carleo}, \citenamefont {Lovato},\ and\ \citenamefont {Rocco}}]{Adams2021Phys.Rev.Lett.022502}%
  \BibitemOpen
  \bibfield  {author} {\bibinfo {author} {\bibfnamefont {C.}~\bibnamefont {Adams}}, \bibinfo {author} {\bibfnamefont {G.}~\bibnamefont {Carleo}}, \bibinfo {author} {\bibfnamefont {A.}~\bibnamefont {Lovato}},\ and\ \bibinfo {author} {\bibfnamefont {N.}~\bibnamefont {Rocco}},\ }\bibfield  {title} {\bibinfo {title} {{Variational Monte Carlo Calculations of $A\ensuremath{\le}4$ Nuclei with an Artificial Neural-Network Correlator Ansatz}},\ }\href {https://doi.org/10.1103/PhysRevLett.127.022502} {\bibfield  {journal} {\bibinfo  {journal} {Phys. Rev. Lett.}\ }\textbf {\bibinfo {volume} {127}},\ \bibinfo {pages} {022502} (\bibinfo {year} {2021})}\BibitemShut {NoStop}%
\bibitem [{\citenamefont {Gnech}\ \emph {et~al.}(2021)\citenamefont {Gnech}, \citenamefont {Adams}, \citenamefont {Brawand}, \citenamefont {Carleo}, \citenamefont {Lovato},\ and\ \citenamefont {Rocco}}]{Gnech2021FewBodySystems7}%
  \BibitemOpen
  \bibfield  {author} {\bibinfo {author} {\bibfnamefont {A.}~\bibnamefont {Gnech}}, \bibinfo {author} {\bibfnamefont {C.}~\bibnamefont {Adams}}, \bibinfo {author} {\bibfnamefont {N.}~\bibnamefont {Brawand}}, \bibinfo {author} {\bibfnamefont {G.}~\bibnamefont {Carleo}}, \bibinfo {author} {\bibfnamefont {A.}~\bibnamefont {Lovato}},\ and\ \bibinfo {author} {\bibfnamefont {N.}~\bibnamefont {Rocco}},\ }\bibfield  {title} {\bibinfo {title} {{Nuclei with Up to $A=6$ Nucleons with Artificial Neural Network Wave Functions}},\ }\href {https://doi.org/10.1007/s00601-021-01706-0} {\bibfield  {journal} {\bibinfo  {journal} {Few-Body Systems}\ }\textbf {\bibinfo {volume} {63}},\ \bibinfo {pages} {7} (\bibinfo {year} {2021})}\BibitemShut {NoStop}%
\bibitem [{\citenamefont {Yang}\ and\ \citenamefont {Zhao}(2022)}]{Yang2022Phys.Lett.B137587}%
  \BibitemOpen
  \bibfield  {author} {\bibinfo {author} {\bibfnamefont {Y.}~\bibnamefont {Yang}}\ and\ \bibinfo {author} {\bibfnamefont {P.}~\bibnamefont {Zhao}},\ }\bibfield  {title} {\bibinfo {title} {{A consistent description of the relativistic effects and three-body interactions in atomic nuclei}},\ }\href {https://doi.org/https://doi.org/10.1016/j.physletb.2022.137587} {\bibfield  {journal} {\bibinfo  {journal} {Phys. Lett. B}\ }\textbf {\bibinfo {volume} {835}},\ \bibinfo {pages} {137587} (\bibinfo {year} {2022})}\BibitemShut {NoStop}%
\bibitem [{\citenamefont {Fore}\ \emph {et~al.}(2023)\citenamefont {Fore}, \citenamefont {Kim}, \citenamefont {Carleo}, \citenamefont {Hjorth-Jensen}, \citenamefont {Lovato},\ and\ \citenamefont {Piarulli}}]{Fore2023Phys.Rev.Res.033062}%
  \BibitemOpen
  \bibfield  {author} {\bibinfo {author} {\bibfnamefont {B.}~\bibnamefont {Fore}}, \bibinfo {author} {\bibfnamefont {J.~M.}\ \bibnamefont {Kim}}, \bibinfo {author} {\bibfnamefont {G.}~\bibnamefont {Carleo}}, \bibinfo {author} {\bibfnamefont {M.}~\bibnamefont {Hjorth-Jensen}}, \bibinfo {author} {\bibfnamefont {A.}~\bibnamefont {Lovato}},\ and\ \bibinfo {author} {\bibfnamefont {M.}~\bibnamefont {Piarulli}},\ }\bibfield  {title} {\bibinfo {title} {{Dilute neutron star matter from neural-network quantum states}},\ }\href {https://doi.org/10.1103/PhysRevResearch.5.033062} {\bibfield  {journal} {\bibinfo  {journal} {Phys. Rev. Res.}\ }\textbf {\bibinfo {volume} {5}},\ \bibinfo {pages} {033062} (\bibinfo {year} {2023})}\BibitemShut {NoStop}%
\bibitem [{\citenamefont {Gnech}\ \emph {et~al.}(2024)\citenamefont {Gnech}, \citenamefont {Fore}, \citenamefont {Tropiano},\ and\ \citenamefont {Lovato}}]{Gnech2024Phys.Rev.Lett.142501}%
  \BibitemOpen
  \bibfield  {author} {\bibinfo {author} {\bibfnamefont {A.}~\bibnamefont {Gnech}}, \bibinfo {author} {\bibfnamefont {B.}~\bibnamefont {Fore}}, \bibinfo {author} {\bibfnamefont {A.~J.}\ \bibnamefont {Tropiano}},\ and\ \bibinfo {author} {\bibfnamefont {A.}~\bibnamefont {Lovato}},\ }\bibfield  {title} {\bibinfo {title} {{Distilling the Essential Elements of Nuclear Binding via Neural-Network Quantum States}},\ }\href {https://doi.org/10.1103/PhysRevLett.133.142501} {\bibfield  {journal} {\bibinfo  {journal} {Phys. Rev. Lett.}\ }\textbf {\bibinfo {volume} {133}},\ \bibinfo {pages} {142501} (\bibinfo {year} {2024})}\BibitemShut {NoStop}%
\bibitem [{\citenamefont {Yang}\ and\ \citenamefont {Zhao}(2025)}]{Yang2025ChinesePhys.Lett.051201}%
  \BibitemOpen
  \bibfield  {author} {\bibinfo {author} {\bibfnamefont {Y.-L.}\ \bibnamefont {Yang}}\ and\ \bibinfo {author} {\bibfnamefont {P.-W.}\ \bibnamefont {Zhao}},\ }\bibfield  {title} {\bibinfo {title} {Reconciling light nuclei and nuclear matter: Relativistic ab initio calculations},\ }\href {https://doi.org/10.1088/0256-307X/42/5/051201} {\bibfield  {journal} {\bibinfo  {journal} {Chinese Phys. Lett.}\ }\textbf {\bibinfo {volume} {42}},\ \bibinfo {pages} {051201} (\bibinfo {year} {2025})}\BibitemShut {NoStop}%
\bibitem [{\citenamefont {Luo}\ and\ \citenamefont {Clark}(2019)}]{Luo2019Phys.Rev.Lett.226401}%
  \BibitemOpen
  \bibfield  {author} {\bibinfo {author} {\bibfnamefont {D.}~\bibnamefont {Luo}}\ and\ \bibinfo {author} {\bibfnamefont {B.~K.}\ \bibnamefont {Clark}},\ }\bibfield  {title} {\bibinfo {title} {{Backflow {Transformations} via {Neural} {Networks} for {Quantum} {Many}-{Body} {Wave} {Functions}}},\ }\href {https://doi.org/10.1103/PhysRevLett.122.226401} {\bibfield  {journal} {\bibinfo  {journal} {Phys. Rev. Lett.}\ }\textbf {\bibinfo {volume} {122}},\ \bibinfo {pages} {226401} (\bibinfo {year} {2019})}\BibitemShut {NoStop}%
\bibitem [{\citenamefont {Yang}\ and\ \citenamefont {Zhao}(2023)}]{Yang2023Phys.Rev.C034320}%
  \BibitemOpen
  \bibfield  {author} {\bibinfo {author} {\bibfnamefont {Y.~L.}\ \bibnamefont {Yang}}\ and\ \bibinfo {author} {\bibfnamefont {P.~W.}\ \bibnamefont {Zhao}},\ }\bibfield  {title} {\bibinfo {title} {{Deep-neural-network approach to solving the ab initio nuclear structure problem}},\ }\href {https://doi.org/10.1103/PhysRevC.107.034320} {\bibfield  {journal} {\bibinfo  {journal} {Phys. Rev. C}\ }\textbf {\bibinfo {volume} {107}},\ \bibinfo {pages} {034320} (\bibinfo {year} {2023})}\BibitemShut {NoStop}%
\bibitem [{\citenamefont {Wiringa}\ \emph {et~al.}(2000)\citenamefont {Wiringa}, \citenamefont {Pieper}, \citenamefont {Carlson},\ and\ \citenamefont {Pandharipande}}]{Wiringa2000Phys.Rev.C014001}%
  \BibitemOpen
  \bibfield  {author} {\bibinfo {author} {\bibfnamefont {R.~B.}\ \bibnamefont {Wiringa}}, \bibinfo {author} {\bibfnamefont {S.~C.}\ \bibnamefont {Pieper}}, \bibinfo {author} {\bibfnamefont {J.}~\bibnamefont {Carlson}},\ and\ \bibinfo {author} {\bibfnamefont {V.~R.}\ \bibnamefont {Pandharipande}},\ }\bibfield  {title} {\bibinfo {title} {{Quantum Monte Carlo calculations of $A=8$ nuclei}},\ }\href {https://doi.org/10.1103/PhysRevC.62.014001} {\bibfield  {journal} {\bibinfo  {journal} {Phys. Rev. C}\ }\textbf {\bibinfo {volume} {62}},\ \bibinfo {pages} {014001} (\bibinfo {year} {2000})}\BibitemShut {NoStop}%
\bibitem [{\citenamefont {Hornik}\ \emph {et~al.}(1989)\citenamefont {Hornik}, \citenamefont {Stinchcombe},\ and\ \citenamefont {White}}]{Hornik1989NeuralNetworks359}%
  \BibitemOpen
  \bibfield  {author} {\bibinfo {author} {\bibfnamefont {K.}~\bibnamefont {Hornik}}, \bibinfo {author} {\bibfnamefont {M.}~\bibnamefont {Stinchcombe}},\ and\ \bibinfo {author} {\bibfnamefont {H.}~\bibnamefont {White}},\ }\bibfield  {title} {\bibinfo {title} {Multilayer feedforward networks are universal approximators},\ }\href {https://doi.org/https://doi.org/10.1016/0893-6080(89)90020-8} {\bibfield  {journal} {\bibinfo  {journal} {Neural Networks}\ }\textbf {\bibinfo {volume} {2}},\ \bibinfo {pages} {359} (\bibinfo {year} {1989})}\BibitemShut {NoStop}%
\bibitem [{\citenamefont {Pudliner}\ \emph {et~al.}(1997)\citenamefont {Pudliner}, \citenamefont {Pandharipande}, \citenamefont {Carlson}, \citenamefont {Pieper},\ and\ \citenamefont {Wiringa}}]{Pudliner1997Phys.Rev.C1720}%
  \BibitemOpen
  \bibfield  {author} {\bibinfo {author} {\bibfnamefont {B.~S.}\ \bibnamefont {Pudliner}}, \bibinfo {author} {\bibfnamefont {V.~R.}\ \bibnamefont {Pandharipande}}, \bibinfo {author} {\bibfnamefont {J.}~\bibnamefont {Carlson}}, \bibinfo {author} {\bibfnamefont {S.~C.}\ \bibnamefont {Pieper}},\ and\ \bibinfo {author} {\bibfnamefont {R.~B.}\ \bibnamefont {Wiringa}},\ }\bibfield  {title} {\bibinfo {title} {{Quantum Monte Carlo calculations of nuclei with $A<7$}},\ }\href {https://doi.org/10.1103/PhysRevC.56.1720} {\bibfield  {journal} {\bibinfo  {journal} {Phys. Rev. C}\ }\textbf {\bibinfo {volume} {56}},\ \bibinfo {pages} {1720} (\bibinfo {year} {1997})}\BibitemShut {NoStop}%
\bibitem [{Sup()}]{Supp}%
  \BibitemOpen
  \href@noop {} {}\bibinfo {note} {See Supplemental Material at [URL], which includes Refs.~\cite{Descouvemont2010Rep.Prog.Phys.036301, Pudliner1995Phys.Rev.Lett.43964399, Furnstahl2015JPG034028}, for details on the BERW formula, GFMC calculations, and uncertainty quantification.}\BibitemShut {Stop}%
\bibitem [{Hal()}]{Hale2025}%
  \BibitemOpen
  \href@noop {} {}\bibinfo {note} {Hale, G. M., private communication (2025).}\BibitemShut {Stop}%
\bibitem [{\citenamefont {Zhang}(2020)}]{Zhang2020Phys.Rev.C051602}%
  \BibitemOpen
  \bibfield  {author} {\bibinfo {author} {\bibfnamefont {X.}~\bibnamefont {Zhang}},\ }\bibfield  {title} {\bibinfo {title} {{Extracting free-space observables from trapped interacting clusters}},\ }\href {https://doi.org/10.1103/PhysRevC.101.051602} {\bibfield  {journal} {\bibinfo  {journal} {Phys. Rev. C}\ }\textbf {\bibinfo {volume} {101}},\ \bibinfo {pages} {051602} (\bibinfo {year} {2020})}\BibitemShut {NoStop}%
\bibitem [{\citenamefont {Luu}\ \emph {et~al.}(2010)\citenamefont {Luu}, \citenamefont {Savage}, \citenamefont {Schwenk},\ and\ \citenamefont {Vary}}]{Luu2010Phys.Rev.C82.034003}%
  \BibitemOpen
  \bibfield  {author} {\bibinfo {author} {\bibfnamefont {T.}~\bibnamefont {Luu}}, \bibinfo {author} {\bibfnamefont {M.~J.}\ \bibnamefont {Savage}}, \bibinfo {author} {\bibfnamefont {A.}~\bibnamefont {Schwenk}},\ and\ \bibinfo {author} {\bibfnamefont {J.~P.}\ \bibnamefont {Vary}},\ }\bibfield  {title} {\bibinfo {title} {Nucleon-nucleon scattering in a harmonic potential},\ }\href {https://doi.org/10.1103/PhysRevC.82.034003} {\bibfield  {journal} {\bibinfo  {journal} {Phys. Rev. C}\ }\textbf {\bibinfo {volume} {82}},\ \bibinfo {pages} {034003} (\bibinfo {year} {2010})}\BibitemShut {NoStop}%
\bibitem [{\citenamefont {Li}\ \emph {et~al.}(2021)\citenamefont {Li}, \citenamefont {Yu}, \citenamefont {Peng}, \citenamefont {Lyu},\ and\ \citenamefont {Long}}]{Li2021Phys.Rev.C104.044001}%
  \BibitemOpen
  \bibfield  {author} {\bibinfo {author} {\bibfnamefont {C.}~\bibnamefont {Li}}, \bibinfo {author} {\bibfnamefont {J.}~\bibnamefont {Yu}}, \bibinfo {author} {\bibfnamefont {R.}~\bibnamefont {Peng}}, \bibinfo {author} {\bibfnamefont {S.}~\bibnamefont {Lyu}},\ and\ \bibinfo {author} {\bibfnamefont {B.}~\bibnamefont {Long}},\ }\bibfield  {title} {\bibinfo {title} {Trapped two-nucleon system in energy-dependent effective field theory},\ }\href {https://doi.org/10.1103/PhysRevC.104.044001} {\bibfield  {journal} {\bibinfo  {journal} {Phys. Rev. C}\ }\textbf {\bibinfo {volume} {104}},\ \bibinfo {pages} {044001} (\bibinfo {year} {2021})}\BibitemShut {NoStop}%
\bibitem [{\citenamefont {Bernard}\ \emph {et~al.}(2008)\citenamefont {Bernard}, \citenamefont {Epelbaum}, \citenamefont {Krebs},\ and\ \citenamefont {Meissner}}]{Bernard:2007sp}%
  \BibitemOpen
  \bibfield  {author} {\bibinfo {author} {\bibfnamefont {V.}~\bibnamefont {Bernard}}, \bibinfo {author} {\bibfnamefont {E.}~\bibnamefont {Epelbaum}}, \bibinfo {author} {\bibfnamefont {H.}~\bibnamefont {Krebs}},\ and\ \bibinfo {author} {\bibfnamefont {U.-G.}\ \bibnamefont {Meissner}},\ }\bibfield  {title} {\bibinfo {title} {{Subleading contributions to the chiral three-nucleon force. I. Long-range terms}},\ }\href {https://doi.org/10.1103/PhysRevC.77.064004} {\bibfield  {journal} {\bibinfo  {journal} {Phys. Rev. C}\ }\textbf {\bibinfo {volume} {77}},\ \bibinfo {pages} {064004} (\bibinfo {year} {2008})}\BibitemShut {NoStop}%
\bibitem [{\citenamefont {Krebs}\ \emph {et~al.}(2012)\citenamefont {Krebs}, \citenamefont {Gasparyan},\ and\ \citenamefont {Epelbaum}}]{Krebs2012Phys.Rev.C054006}%
  \BibitemOpen
  \bibfield  {author} {\bibinfo {author} {\bibfnamefont {H.}~\bibnamefont {Krebs}}, \bibinfo {author} {\bibfnamefont {A.}~\bibnamefont {Gasparyan}},\ and\ \bibinfo {author} {\bibfnamefont {E.}~\bibnamefont {Epelbaum}},\ }\bibfield  {title} {\bibinfo {title} {{Chiral three-nucleon force at N${}^{4}$LO: Longest-range contributions}},\ }\href {https://doi.org/10.1103/PhysRevC.85.054006} {\bibfield  {journal} {\bibinfo  {journal} {Phys. Rev. C}\ }\textbf {\bibinfo {volume} {85}},\ \bibinfo {pages} {054006} (\bibinfo {year} {2012})}\BibitemShut {NoStop}%
\bibitem [{\citenamefont {Epelbaum}\ \emph {et~al.}(2020)\citenamefont {Epelbaum}, \citenamefont {Golak}, \citenamefont {Hebeler}, \citenamefont {Kamada}, \citenamefont {Krebs}, \citenamefont {Mei\ss{}ner}, \citenamefont {Nogga}, \citenamefont {Reinert}, \citenamefont {Skibi\'nski}, \citenamefont {Topolnicki}, \citenamefont {Volkotrub},\ and\ \citenamefont {Wita\l~a}}]{Epelbaum2020EPJA92}%
  \BibitemOpen
  \bibfield  {author} {\bibinfo {author} {\bibfnamefont {E.}~\bibnamefont {Epelbaum}}, \bibinfo {author} {\bibfnamefont {J.}~\bibnamefont {Golak}}, \bibinfo {author} {\bibfnamefont {K.}~\bibnamefont {Hebeler}}, \bibinfo {author} {\bibfnamefont {H.}~\bibnamefont {Kamada}}, \bibinfo {author} {\bibfnamefont {H.}~\bibnamefont {Krebs}}, \bibinfo {author} {\bibfnamefont {U.-G.}\ \bibnamefont {Mei\ss{}ner}}, \bibinfo {author} {\bibfnamefont {A.}~\bibnamefont {Nogga}}, \bibinfo {author} {\bibfnamefont {P.}~\bibnamefont {Reinert}}, \bibinfo {author} {\bibfnamefont {R.}~\bibnamefont {Skibi\'nski}}, \bibinfo {author} {\bibfnamefont {K.}~\bibnamefont {Topolnicki}}, \bibinfo {author} {\bibfnamefont {Y.}~\bibnamefont {Volkotrub}},\ and\ \bibinfo {author} {\bibfnamefont {H.}~\bibnamefont {Wita\l~a}},\ }\bibfield  {title} {\bibinfo {title} {Towards high-order calculations of three-nucleon scattering in chiral effective field theory},\ }\href {https://doi.org/10.1140/epja/s10050-020-00102-2} {\bibfield  {journal} {\bibinfo
  {journal} {Eur. Phys. J. A}\ }\textbf {\bibinfo {volume} {56}},\ \bibinfo {pages} {92} (\bibinfo {year} {2020})}\BibitemShut {NoStop}%
\bibitem [{\citenamefont {Descouvemont}\ and\ \citenamefont {Baye}(2010)}]{Descouvemont2010Rep.Prog.Phys.036301}%
  \BibitemOpen
  \bibfield  {author} {\bibinfo {author} {\bibfnamefont {P.}~\bibnamefont {Descouvemont}}\ and\ \bibinfo {author} {\bibfnamefont {D.}~\bibnamefont {Baye}},\ }\bibfield  {title} {\bibinfo {title} {{The R-matrix theory}},\ }\href {https://doi.org/10.1088/0034-4885/73/3/036301} {\bibfield  {journal} {\bibinfo  {journal} {Rep. Prog. Phys.}\ }\textbf {\bibinfo {volume} {73}},\ \bibinfo {pages} {036301} (\bibinfo {year} {2010})}\BibitemShut {NoStop}%
\bibitem [{\citenamefont {Pudliner}\ \emph {et~al.}(1995)\citenamefont {Pudliner}, \citenamefont {Pandharipande}, \citenamefont {Carlson},\ and\ \citenamefont {Wiringa}}]{Pudliner1995Phys.Rev.Lett.43964399}%
  \BibitemOpen
  \bibfield  {author} {\bibinfo {author} {\bibfnamefont {B.~S.}\ \bibnamefont {Pudliner}}, \bibinfo {author} {\bibfnamefont {V.~R.}\ \bibnamefont {Pandharipande}}, \bibinfo {author} {\bibfnamefont {J.}~\bibnamefont {Carlson}},\ and\ \bibinfo {author} {\bibfnamefont {R.~B.}\ \bibnamefont {Wiringa}},\ }\bibfield  {title} {\bibinfo {title} {{Quantum Monte Carlo Calculations of $A\leq6$ Nuclei}},\ }\href {https://doi.org/10.1103/PhysRevLett.74.4396} {\bibfield  {journal} {\bibinfo  {journal} {Phys. Rev. Lett.}\ }\textbf {\bibinfo {volume} {74}},\ \bibinfo {pages} {4396} (\bibinfo {year} {1995})}\BibitemShut {NoStop}%
\bibitem [{\citenamefont {Furnstahl}\ \emph {et~al.}(2015)\citenamefont {Furnstahl}, \citenamefont {Phillips},\ and\ \citenamefont {Wesolowski}}]{Furnstahl2015JPG034028}%
  \BibitemOpen
  \bibfield  {author} {\bibinfo {author} {\bibfnamefont {R.~J.}\ \bibnamefont {Furnstahl}}, \bibinfo {author} {\bibfnamefont {D.~R.}\ \bibnamefont {Phillips}},\ and\ \bibinfo {author} {\bibfnamefont {S.}~\bibnamefont {Wesolowski}},\ }\bibfield  {title} {\bibinfo {title} {A recipe for eft uncertainty quantification in nuclear physics},\ }\href {https://doi.org/10.1088/0954-3899/42/3/034028} {\bibfield  {journal} {\bibinfo  {journal} {J. Phys. G}\ }\textbf {\bibinfo {volume} {42}},\ \bibinfo {pages} {034028} (\bibinfo {year} {2015})}\BibitemShut {NoStop}%
\end{thebibliography}

%

\end{document}